\documentclass[lettersize,journal]{IEEEtran}
\usepackage{amsmath,amsfonts}
\usepackage{algorithmic}
\usepackage{algorithm}
\usepackage{array}
\usepackage[caption=false,font=normalsize,labelfont=sf,textfont=sf]{subfig}
\usepackage{textcomp}
\usepackage{stfloats}
\usepackage{url}
\usepackage{verbatim}
\usepackage{graphicx}
\usepackage{cite}
\usepackage{pifont}  %
\usepackage{xcolor}  %
\definecolor{bedroom_color}{RGB}{251,154,153}
\definecolor{kitchen_color}{RGB}{178,223,128}
\newcommand{\cmark}{{\color{kitchen_color}\ding{51}}}  %
\newcommand{\xmark}{{\color{bedroom_color}\ding{55}}}  %

\usepackage{multirow}
\usepackage{tikz}

\begin{document}

\title{Directly from Alpha to Omega: \\Controllable End-to-End Vector Floor Plan Generation}

\author{Shidong~Wang and Renato~Pajarola,~\IEEEmembership{Senior Member,~IEEE}
\thanks{S.~Wang and R.~Pajarola are with the Department of Informatics, University of Zurich, Switzerland. E-mails: \{shwang,~pajarola\}@ifi.uzh.ch.}
\thanks{Received 13 September 2025; revised 7 January 2026; accepted 7 February 2026.}}

\markboth{IEEE Transactions on Visualization and Computer Graphics}%
{Wang \MakeLowercase{\textit{et al.}}: Directly from Alpha to Omega: Controllable End-to-End Vector Floor Plan Generation}


\maketitle


\begin{abstract}

Automated floor plan generation aims to create residential layouts by arranging rooms within a given boundary, balancing topological, geometric, and aesthetic considerations.
The existing methods typically use a multi-step pipeline with intermediate representations to decompose the prediction process into several sub-tasks, limiting model flexibility and imposing predefined solution paths.
This often results in unreasonable outputs when applied to data unsuitable for these predefined paths, making it challenging for these methods to match human designers, who do not restrict themselves to a specific set of design workflows.
To address these limitations, we introduce CE2EPlan, a controllable end-to-end topology- and geometry-enhanced diffusion model that removes restrictions on the generative process of AI design tools.
Instead, it enables the model to learn how to design floor plans directly from data, capturing a wide range of solution paths from input boundaries to complete layouts.
Extensive experiments demonstrate that our method surpasses all existing approaches using the multi-step pipeline, delivering higher-quality results with enhanced user control and greater diversity in output, bringing AI design tools closer to the versatility of human designers.

\end{abstract}


\section{Introduction}  \label{sec:introduction}

Automated floor plan generation aims to create an arrangement of rooms, including their geometric and topological properties, within a given exterior wall boundary of the residential building, which has drawn significant research interest over the past decades in computer graphics~\cite{Wu-19, Hu-20, Sun-22}, vision~\cite{Nauata-20, Nauata-21, He-22}, and architecture~\cite{Arvin-02, Michalek-02, Rodrigues-13}.
Conventional methods~\cite{Merrell-10, Bao-13, Wu-18} rely on design constraints and optimization, but require precise constraint settings.
Too few yield poor results, too many lead to contradictions, and some constraints are challenging to model mathematically~\cite{Sun-22}.

\begin{figure*}[t]
    \centering
    \includegraphics[width=\linewidth]{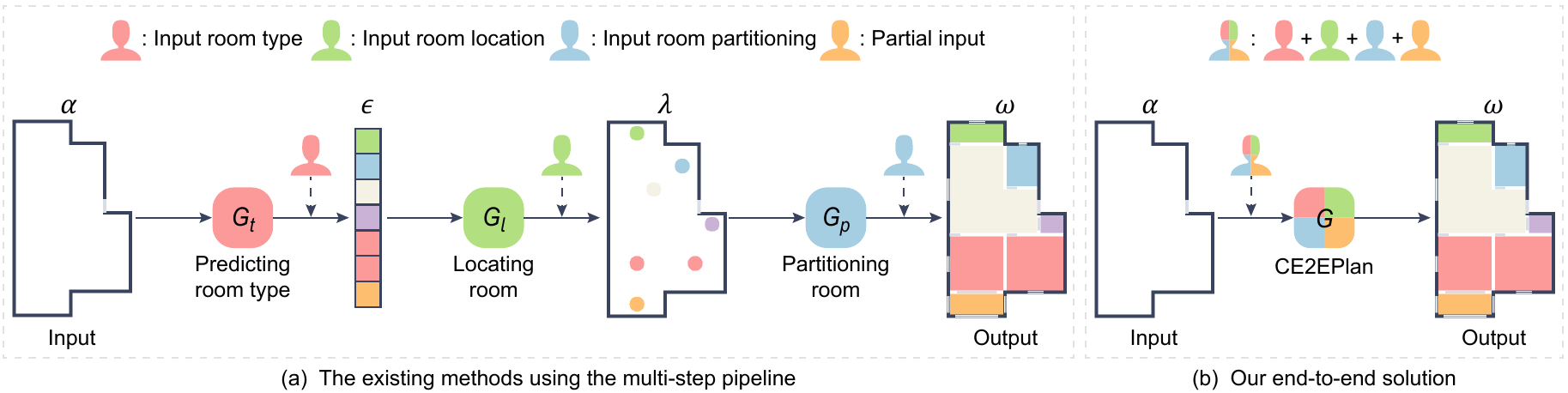}
    \caption{Given a boundary as input, unlike previous methods (a) that decompose the design process into several sub-tasks with fixed solution paths ($\alpha\rightarrow\epsilon\rightarrow\lambda\rightarrow\omega$), our CE2EPlan (b) learns floor plan creation directly from data without predefined pathways. It captures diverse design paths from the input boundary to the output floor plan ($\alpha\rightarrow\omega$) and delivers higher-quality results with enhanced user control and output diversity.}
    \label{fig:teaser}
\end{figure*}

Therefore, learning-based methods have been proposed to avoid manually defined constraints and instead implicitly learn the design principles from existing data, enabling more efficient floor plan generation.
Current learning-based methods~\cite{Wu-19, Chaillou-20, Hu-20, Wang-21, He-22, Sun-22, Wang-25} often break down the prediction process of floor plans into several sub-processes, using a multi-step pipeline with intermediate representations to handle floor plan generation.
Figure~\ref{fig:teaser}(a) shows a typical breakdown process: Starting from a given boundary, these methods first predict room types, then room locations, and finally room shapes and partitioning ($\alpha\rightarrow\epsilon\rightarrow\lambda\rightarrow\omega$).

There are two main reasons for adopting this multi-step pipeline.
First, it was adopted to improve the quality of the generated results.
Floor plan generation is a complex task involving topology, geometry, and aesthetics~\cite{Wang-21}.
Previously, directly predicting the floor plan from the input boundary yielded unsatisfactory results~\cite{Chaillou-20}.
Therefore, past methods chose to break down the problem into several sub-tasks, designing specific models for each sub-task to improve the plausibility of the final generated results.
Second, it offers controllability of the model and output diversity.
Users can adjust the output of the preceding models to influence subsequent predictions, achieving good control.
Diversity in the results can also be obtained by making slight adjustments to the output of the preceding models.
Nevertheless, as we demonstrate, better results can be achieved without following such a multi-step concept and instead using an end-to-end strategy as illustrated in Figure~\ref{fig:teaser}(b).

The original intent of learning-based methods is to minimize manually defined design constraints~\cite{Wu-19, Wang-21}.
However, structuring the floor plan generation process according to a hypothesized human design workflow already diverges from this initial goal, imposing design constraints at a more macro level than conventional methods.
These approaches using the multi-step pipeline restrict the deductive process of AI design tools, e.g., always predicting types \& locations before partitioning~\cite{Wu-19, He-22}, or deriving layouts from bubble diagrams~\cite{Hu-20, Shabani-23, Wang-25}, human-activity maps~\cite{Wang-21}, or other specific conditions~\cite{Chaillou-20, Sun-22}.
Limiting an AI design tool to one or a few fixed solution paths for solving design problems inherently reduces its ability to match the versatility of human designers.

We thus propose a new approach -- one that, from a methodological perspective, does not rely on a predefined solution path with intermediate representations.
Instead, our goal is an end-to-end model that directly infers floor plans from the given boundary after learning from large-scale data ($\alpha\rightarrow\omega$).
Functionally, it should provide enhanced controllability by allowing users to specify input conditions (e.g., room types and locations) and enabling iterative user interaction, and support true output diversity, meaning a single model generating multiple distinct outputs for the same input, rather than only relying on minor modifications to the output of preceding models to achieve variation.
This task is evidently challenging, much like an old Chinese proverb: ``You want the horse to run fast (meaning that the model should generate high-quality results while possessing rich controllability and output diversity), but you do not want to feed it grass (meaning that the model is expected to learn these abilities directly from data in an end-to-end manner, without being provided predefined solution paths, using intermediate representations to break the task into sub-tasks to reduce complexity)''.

Based on this, we propose CE2EPlan, a controllable topology- and geometry-enhanced diffusion model for end-to-end floor plan generation.
Specifically, to capture the high-dimensional design space that involves topology, geometry, and aesthetics in generating floor plans directly from an input boundary, it utilizes a recent powerful generative approach, the diffusion model (DM)~\cite{Ho-20}, as its backbone.
A multi-condition masking mechanism enables the controllability of the model, while a GATransformer, an integration of Transformer~\cite{Vaswani-17} and Graph Attention Networks (GATs)~\cite{Velickovic-18-GAT}, is proposed as the noise predictor of the DM.
This setup allows the model to handle complex vector floor plan data, extract topological features, and improve the topological coherence of the predicted results.
Several loss functions focused on floor plan alignment are integrated into the training of the model to enhance geometric consistency in the final output.
Due to the reverse process of DM starting from a standard Gaussian random variable $x_T \sim \mathcal{N}(0,\mathbf{I})$, we can easily obtain diverse outputs for the same input.
Additionally, by simply disabling the input boundary during the training and testing, boundary-unconstrained floor plan generation can also be achieved.

In summary, the contributions of our CE2EPlan approach are as follows:

\begin{enumerate}

\item 
We introduce the first end-to-end model for floor plan generation that eliminates the need for intermediate representations and predefined solution paths, learning directly from existing data to predict floor plans from the input.

\item
All interaction modes and genuine generative diversity are supported within a single unified model that requires only one training process.
This stands in contrast to prior multi-step pipelines that rely on separately trained sub-models for different input conditions and variable outputs from preceding models in the sequential pipeline.

\item
We design a simple yet expressive data representation, a multi-condition masking mechanism, and topology- and geometry-enhanced modules that collectively make the above capabilities feasible.

\end{enumerate}

\begin{table*}[t]
    \centering
    \caption{Compared to the state-of-the-art methods for floor plan generation, our CE2EPlan differs in both methodology and functionality. From a methodological perspective, CE2EPlan does not rely on a multi-step pipeline with intermediate representations but instead directly predicts the layout end-to-end from the input ($\alpha\rightarrow\omega$). From a functional perspective, CE2EPlan requires only a single training process to support richer interaction modes, i.e., without user input ($\text{Mode}_{\text{auto}}$), using room type input ($\text{Mode}_{\text{t}}$), using both room type \& location input ($\text{Mode}_{\text{t}\&\text{l}}$), and partial input ($\text{Mode}_{\text{part}}$), and achieve true output diversity, with a single model itself producing multiple predictions for the same input rather than relying on varied outputs from preceding models. $*$ indicates that additional room adjacencies are required as the input.}
    \label{tab:overview}
    \begin{tabular}{|l|cc|cccc|c|}
        \hline
        & \multicolumn{2}{c|}{Methodology} & \multicolumn{5}{c|}{Functionality}  \\
        \cline{2-8}
        & \multicolumn{2}{c|}{Solution Path} & \multicolumn{4}{c|}{Mode / Number of Required Models} & Output \\
        Method & End-to-End  & Multi-step Pipeline & $\text{Mode}_{\text{auto}}$ & $\text{Mode}_{\text{t}}$  & $\text{Mode}_{\text{t}\&\text{l}}$ &  $\text{Mode}_{\text{part}}$ & Diversity \\
        \hline
        RPLAN~\cite{Wu-19} & \xmark & 
        \begin{tikzpicture}[baseline=(base.base)]
				\node[inner sep=0] (base) at (0,0) {\strut};
				\node (A) at (0,0) {$\alpha$};
				\node (B) at (0.8,0) {$\delta$};
				\node (C) at (1.6,0) {$\lambda$};
				\node (D) at (2.4,0) {$\sigma$};
				\node (E) at (3.2,0) {$\omega$};
				\draw[->] (A) -- (B);
				\draw[->] (B) -- (C);
				\draw[->] (C) -- (D);
				\draw[->] (D) -- (E);
				\draw[->, bend right=20] (D) to (B);
			\end{tikzpicture} & \cmark / 4 & \xmark & \cmark / 1  & \xmark & \xmark \\
        ActFloor-GAN~\cite{Wang-21}  & \xmark & 
        \begin{tikzpicture}[baseline=(base.base)]
				\node[inner sep=0] (base) at (0,0) {\strut};
				\node (A) at (0,0) {$\alpha$};
				\node (B) at (0.8,0) {$\theta$};
				\node (C) at (1.6,0) {$\omega$};
				\draw[->] (A) -- (B);
				\draw[->] (B) -- (C);
			\end{tikzpicture} & \cmark / 2 & \xmark & \xmark & \xmark & \xmark \\
        WallPlan~\cite{Sun-22}  & \xmark & 
        \begin{tikzpicture}[baseline=(base.base)]
				\node[inner sep=0] (base) at (0,0) {\strut};
				\node (A) at (0,0) {$\alpha$};
				\node (B) at (0.8,0) {$\gamma$};
				\node (C) at (1.6,0) {$\eta$};
				\node (D) at (2.4,0) {$\rho$};
				\node (E) at (3.2,0) {$\omega$};
				\draw[->] (A) -- (B);
				\draw[->] (B) -- (C);
				\draw[->] (C) -- (D);
				\draw[->] (D) -- (E);
				\draw[->, bend right=20] (E) to (C);
			\end{tikzpicture} & \cmark / 4 & \xmark & \xmark & \xmark & \xmark \\
        Graph2Plan~\cite{Hu-20}  & \xmark & 
        \begin{tikzpicture}[baseline=(base.base)]
				\node[inner sep=0] (base) at (0,0) {\strut};
				\node (A) at (0,0) {$\alpha$};
				\node (B) at (0.8,0) {$\lambda*$};
				\node (C) at (1.6,0) {$\omega$};
				\draw[->] (A) -- (B);
				\draw[->] (B) -- (C);
			\end{tikzpicture} & \cmark / 2 & \xmark & \cmark$*$ / 1 & \xmark & \xmark \\
        DiffPlanner~\cite{Wang-25}  & \xmark & 
        \begin{tikzpicture}[baseline=(base.base)]
				\node[inner sep=0] (base) at (0,0) {\strut};
				\node (A) at (0,0) {$\alpha$};
				\node (B) at (0.8,0) {$\lambda$};
				\node (C) at (1.6,0) {$\lambda*$};
				\node (D) at (2.4,0) {$\omega$};
				\draw[->] (A) -- (B);
				\draw[->] (B) -- (C);
				\draw[->] (C) -- (D);
			\end{tikzpicture} & \cmark / 3 & \cmark / 3 & \cmark / 2 & \xmark & \xmark \\
        iPLAN~\cite{He-22}  & \xmark & 
        \begin{tikzpicture}[baseline=(base.base)]
				\node[inner sep=0] (base) at (0,0) {\strut};
				\node (A) at (0,0) {$\alpha$};
				\node (B) at (0.8,0) {$\epsilon$};
				\node (C) at (1.6,0) {$\lambda$};
				\node (D) at (2.4,0) {$\omega$};
				\draw[->] (A) -- (B);
				\draw[->] (B) -- (C);
				\draw[->] (C) -- (D);
				\draw[->, bend right=20] (C) to (B);
				\draw[->, bend right=20] (D) to (C);
			\end{tikzpicture} & \cmark / 3 & \cmark / 2 & \cmark / 1 & \xmark & \xmark \\
        HouseDiffusion~\cite{Shabani-23}  & 
        \xmark & \begin{tikzpicture}[baseline=(base.base)]
				\node[inner sep=0] (base) at (0,0) {\strut};
				\node (A) at (0,0) {$\epsilon*$};
				\node (B) at (0.8,0) {$\omega$};
				\draw[->] (A) -- (B);
			\end{tikzpicture} & \xmark & \cmark$*$ / 1 & \xmark & \xmark & \cmark \\
        CE2EPlan [Our] & 
        \begin{tikzpicture}[baseline=(base.base)]
				\node[inner sep=0] (base) at (0,0) {\strut};
				\node (A) at (0,0) {$\alpha$};
				\node (B) at (0.8,0) {$\omega$};
				\draw[->] (A) -- (B);
			\end{tikzpicture}  & \xmark & \cmark / 1 & \cmark / 1 & \cmark / 1 & \cmark / 1 & \cmark \\
        \hline
    \end{tabular}  
\end{table*}

\section{Related Work}  \label{sec:related_work}

\subsection{Optimization-based floor plan generation} \label{subsec:rw_optimization}

Conventional methods~\cite{Merrell-10, Bao-13, Wu-18} for automated floor plan generation typically adopt a two-stage optimization-based approach, first determining the relevant constraints, and then optimizing the layout.
However, designing floor plans involves more than geometric and topological considerations; it also requires attention to aesthetics and usability, aspects that are harder to quantify~\cite{Wang-21}.
Although common geometric and topological properties can be represented through objectives like cost and performance, aesthetic and practical layout concerns often defy straightforward modeling.
Furthermore, balancing constraints poses a significant challenge.
Too few constraints lead to suboptimal designs, while too many can create conflicts, making it difficult to find viable solutions~\cite{Sun-22}.

\subsection{Learning-based floor plan generation} \label{subsec:rw_learning}

In recent years, learning-based methods~\cite{Wu-19, Chaillou-20, Hu-20, Nauata-20, Nauata-21, Wang-21, He-22, Sun-22, Shabani-23, Wang-25} for floor plan generation have gained popularity by bypassing the need for manually defined constraints in conventional methods, instead learning design principles directly from data.
However, the existing methods often divide the prediction task into several sub-steps, using multi-step pipelines to handle the complexities of floor plan generation.
As shown in Table~\ref{tab:overview}, previous methods require at least one intermediate step, first transforming the input ($\alpha$) into the intermediate representations, e.g., room types ($\epsilon$), room locations (living room: $\delta$, all rooms: $\lambda$), room windows (living room: $\gamma$, all rooms: $\eta$), human-activity maps ($\theta$), or bubble diagrams (w/o Locations: $\epsilon*$, w/ Locations: $\lambda*$), before converting them into the final floor plans ($\omega$).

Another work, PlanNet~\cite{fu2024plannet}, adopts a staged design workflow for component-based plan synthesis, and recent LLM-based approaches~\cite{qiu2025llmbased, zong2024housetune} generate floor plans from textual descriptions.
These methods differ substantially from the approaches summarized in Table~\ref{tab:overview} in terms of input modalities and problem formulation.
Nevertheless, similar to prior methods, they still follow multi-stage generation pipelines.
Adopting the multi-step pipeline aims to improve the output quality and user control of the model, but it reintroduces manual intervention by defining a fixed solution path from input to the output floor plans, limiting model flexibility.
Human designers, however, do not strictly adhere to predefined solution paths, suggesting that an AI design tool based on a multi-step pipeline may struggle to match the adaptability of human designers.

In contrast, our method is a controllable end-to-end model that directly generates the final floor plans from the input, without any intermediary steps ($\alpha\rightarrow\omega$).
Despite being an end-to-end model, it retains rich controllability, allowing for user-defined conditions and interactive iteration after a single training process.
Our CE2EPlan also achieves true output diversity by generating varying outputs for the same input, whereas the previous methods can only produce different results by altering the outputs of preceding models in the pipeline.

\subsection{Diffusion models (DM)} \label{subsec:rw_dm}

Diffusion models (DM)~\cite{Ho-20, song2020ddim} are a class of emerging generative models that utilize a reverse noise addition process to gradually generate realistic data from random noise.
In practice, DM have been successfully applied in many fields, including natural language processing~\cite{li2022diffusion, strudel2022self}, text-to-image conversion~\cite{ramesh2022hierarchical, zhang2023text}, point cloud generation~\cite{luo2021diffusion, lyu2021conditional}, as well as various layout generation tasks~\cite{inoue2023layoutdm, Shabani-23, chen2024lace, Wang-25}.

In document and UI layout synthesis, LayoutDM~\cite{inoue2023layoutdm} represents layout elements as discrete tokens and applies a discrete diffusion process for controllable arrangement, while LACE~\cite{chen2024lace} incorporates aesthetic constraints to refine the placement of UI components.
These works demonstrate the capability of DM for structured 2D arrangement tasks.
However, the nature of their target domains differs substantially from architectural layout generation.
Document and UI layouts consist of independent visual or functional elements that can be arranged relatively freely, typically without being constrained by a global enclosing boundary.
In contrast, architectural floor plans involve functionally meaningful spaces (rooms) that must be organized within a fixed building boundary and exhibit domain-specific relationships such as functional grouping (e.g., wet areas), separation (e.g., private vs. public zones), and hierarchical circulation.
As a result, floor plan generation poses a more structured spatial reasoning problem, where the arrangement of rooms defines not only visual layout but also the functional usability of the resulting space.

Recently, HouseDiffusion~\cite{Shabani-23} uses DM to predict floor plans from bubble diagrams that encode room counts, types, and adjacencies but exclude boundaries.
Same as the prior work~\cite{Nauata-20, Nauata-21}, it employs the bubble diagram as an intermediate representation, dividing floor plan generation into two sub-tasks and predetermining a solution path ($\alpha\dashrightarrow\epsilon*\rightarrow\omega$).
Fundamentally, it still follows a multi-step pipeline approach.
However, these methods focus only on the later sub-task ($\epsilon*\rightarrow\omega$) -- converting bubble diagrams, which represent the well-defined design intent of architects, into final floor plans.
The crucial earlier sub-task ($\alpha\rightarrow\epsilon*$), which pertains to the early creative phase of the design process, remains unaddressed.
A more recent approach, DiffPlanner~\cite{Wang-25}, proposes a multi-step pipeline based on DM to generate floor plans either from a given boundary or from scratch, first predicting room nodes, then room adjacencies, and finally performing room partitioning ($\alpha\rightarrow\lambda\rightarrow\lambda*\rightarrow\omega$).

In contrast to DiffPlanner, our new method aims to empower the model to function as a true AI designer.
Instead of breaking down the design process with intermediate representations, our CE2EPlan learns to generate floor plans either from a given boundary or from scratch in an end-to-end manner directly from large-scale data ($\alpha\rightarrow\omega$).
Furthermore, our CE2EPlan supports user-defined control conditions and allows iterative interaction, making it a more interactive and flexible tool.

\begin{figure*}[t]
    \centering
    \includegraphics[width=\linewidth]{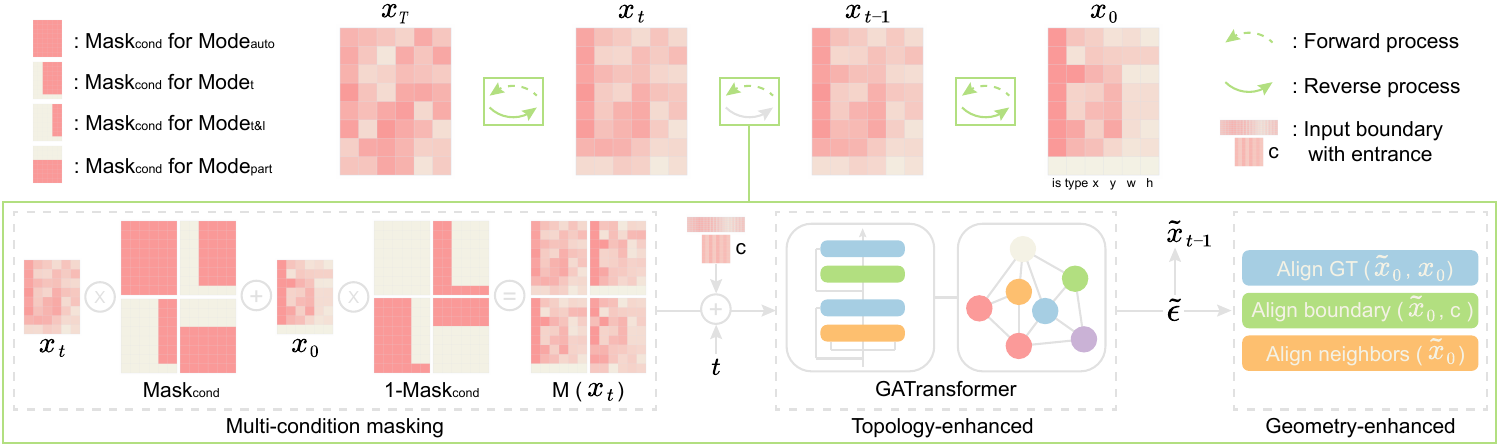}
    \caption{Overview of our CE2EPlan, which adds noise $\epsilon$ to the ground truth data $x_0$ and learns to reverse this process. At each time step $t$, a multi-condition masking mechanism is first used to obtain $M(x_t)$ to support multiple user inputs. Then, the input boundary $c$, $M(x_t)$, \& $t$ are fed into the GATransformer to capture topological properties within the floor plan and infer the corresponding noise $\tilde \epsilon$ and $\tilde x_0$. The predicted $\tilde x_0$ is further refined with three alignment losses to enhance geometric consistency.}
    \label{fig:overview}
\end{figure*}


\section{Method} \label{sec:method}

In this section, we first introduce the data representation used in CE2EPlan.
We then describe its backbone architecture, followed by the multi-condition masking mechanism, which enables controllability.
Next, we present the topology- and geometry-enhanced modules, designed to improve the topological and geometric properties of the generated floor plans.
An overview of our CE2EPlan is presented in Figure~\ref{fig:overview}.

\subsection{Data representation} \label{subsec:method_data}

This work utilizes the RPLAN dataset~\cite{Wu-19} containing $80K$ real-world floor plans, then processes it using the following method into the input and output tensors that our vector-based model can accept, and further randomly splits it into $56K$-$12K$-$12K$ for training-validation-test sets.

\paragraph{Input}
The input exterior wall boundary, including an entrance, is an important condition for typical floor plan generation tasks~\cite{Wu-19}, typically represented by a set of corner coordinates.
Inspired by \cite{Wang-25}, we directly feed the vector boundary and entrance into our model.
The entrance has a fixed number of four corners, and the number of boundary corners is variable, with an upper limit of $40$.
To handle smaller corner lists, we iteratively add a corner at the midpoint of the longest edge of the boundary until the number of corners reaches $40$, which is sufficient to represent the exterior boundary of almost all real-world buildings.
Next, we replicate the boundary and entrance $8$ times, corresponding to the maximum number of rooms in the RPLAN dataset~\cite{Wu-19}.
Finally, we convert the vector boundary and entrance into tensors with dimensions $8\times80$ and $8\times8$, respectively.
Note that for the boundary-unconstrained generation task, we disable the input boundary and entrance.

\paragraph{Output}
We also represent the output rooms as an $8\times6$ tensor in vector space.
Each row of the output tensor represents information for a single room, and the six columns indicate is-room ($1$ for room, $0$ for padding of unused rows), room type ($1$ to $6$ representing living room, bedroom, kitchen, bathroom, balcony, and storage), and the room's bounding box ($x$, $y$, $w$, $h$ as the center coordinates, width, and height).
This representation ensures that the entire prediction process occurs within the vector space, allowing the model to fully capture the topological and geometric properties in the floor plan without discretization to raster images.
Ultimately, we normalize the data tensors of the input and output to a range of $[-1, 1]$.

\subsection{Backbone} \label{subsec:method_backbone}

For capturing intricate patterns and dependencies in vector floor plan data, our CE2EPlan utilizes diffusion models (DM)~\cite{Ho-20}, currently among the most popular generative models, as the backbone.
The DM transform a Gaussian noise $x_T$ into a data sample $x_0$ through a series of $T$ denoising steps, involving both forward and reverse processes during training.
In the forward process $q(x_t | x_0)$, a data sample $x_0$ is progressively converted into a noisy sample $x_t$ at each time step $t$ by adding a Gaussian noise $\epsilon \sim \mathcal{N}(0,\mathbf{I})$:
\begin{equation}
    x_t = \sqrt{\bar \alpha_t}x_0 + \sqrt{(1-\bar \alpha_t)}\epsilon
    \label{eq:forward}
\end{equation}

Here, $\alpha_t = (1 - \beta_t)$, where $\beta_t$ is the variance schedule controlling the noise level, and $\bar{\alpha}_t = \prod_{s=1}^t \alpha_s$ represents the cumulative data preservation.
The reverse process $p_{\theta}(x_{t-1} | x_{t})$ begins with pure Gaussian noise $x_T \sim \mathcal{N}(0,\mathbf{I})$, and iteratively denoises it step by step until it reaches $x_0$.
During this process, the model takes $x_t$ and estimates $x_{t-1}$ by inferring $\epsilon$~\cite{Ho-20}:

\begin{equation}
    x_{t-1} = \frac{1}{\sqrt{\alpha_t}}(x_t - \frac{1-\alpha_t}{\sqrt{1-\bar \alpha_t}}\epsilon_\theta(x_t, t))+\sigma_t z
    \label{eq:reverse}
\end{equation}
where ${\epsilon}_{\theta}$ is a function approximator and is optimized by:

\begin{equation}
    \mathcal{L}_{\mathrm{t-1}} = || {\epsilon}_{\theta} (x_t, t) - {\epsilon} ||^2
    \label{eq:diff_loss}
\end{equation}

\subsection{Multi-condition masking mechanism} \label{subsec:method_masking_module}

\paragraph{Input boundary}
We directly provide the tensors representing the input boundary and entrance as a condition $c$ to the noise predictor $\epsilon_\theta(x_t, t, c)$, ensuring that the boundary information is included throughout the entire diffusion process.
Also, by simply disabling the $c$ in $\epsilon_\theta(x_t, t, c)$ during the training and testing, our method can be applied to boundary-unconstrained generation tasks.

\paragraph{Various user inputs}
User controllability is essential for an AI design tool~\cite{Shneiderman-20}.
Prior methods~\cite{Wu-19, Hu-20, Wang-21, He-22, Sun-22, Wang-25} have primarily achieved controllability by breaking down the design process into multiple sub-tasks, allowing users to modify the outputs of preceding models to influence the outputs of subsequent ones.
In contrast, our goal is for the floor plan generation process to be handled by an end-to-end model.
A straightforward approach is to modify the condition $c$ in the noise predictor $\epsilon_\theta(x_t, t, c)$, training a specific model for each user input condition.
However, this would result in multiple models tailored for different user inputs, which we consider redundant.

To address this, we introduce a masking mechanism to achieve multi-condition controllability.
We classify user input conditions into four categories, i.e., fully automatic generation (without any user input, $\text{Mode}_{\text{auto}}$), specifying room types ($\text{Mode}_{\text{t}}$), specifying both room types and their locations ($\text{Mode}_{\text{t}\&\text{l}}$), and providing partial inputs ($\text{Mode}_{\text{part}}$) to complete the floor plan which is the key to the iterative interaction.
For these four types of input conditions, we generate four corresponding masks ($\text{Mask}_{\text{cond}}$), as shown in Figure~\ref{fig:overview}, which are applied to: $M(x_t) = x_t \cdot \text{Mask}_{\text{cond}} + x_0 \cdot (1 - \text{Mask}_{\text{cond}})$, where $1$ denotes an all-ones matrix.
Then, we feed $M(x_t)$ into the noise predictor $\epsilon_\theta(M(x_t), t, c)$ which can be optimized by: 

\begin{equation}
    \mathcal{L}_{\mathrm{t-1}} = ||{\epsilon}_{\theta} (M(x_t), t, c) - {\epsilon} ||^2 \cdot \text{Mask}_{\text{cond}}
    \label{eq:diff_loss_c}
\end{equation}
Ultimately, our model can be trained end-to-end, requiring only a single training process to support a wide range of user interactions.

\subsection{Topology-enhanced mechanism} \label{subsec:method_topology_module}

The noise predictor $\epsilon_\theta(x_t, t, c)$ is a core component of the DM, and its performance directly impacts the quality of the generated results.
It needs to accurately infer the noise required to reconstruct $x_{t-1}$ from the inputs $x_t$, $t$, \& $c$.
The Transformer~\cite{Vaswani-17} replaced the original U-Net~\cite{Ronneberger-15} as the noise predictor of DM to handle the complex vector layout data in prior work~\cite{Shabani-23, Wang-25}.
However, floor plan generation is often regarded as a graph generation problem~\cite{Hu-20} due to the highly interconnected topological relationships between room nodes in the floor plan.
The model needs to capture these topological relationships between room nodes from the training data to generate reasonable results during testing.

Therefore, we propose GATransformer as the noise predictor, which combines both Transformer encoders~\cite{Vaswani-17} and Graph Attention Networks (GATs)~\cite{Velickovic-18-GAT}, allowing the model to handle complex vector floor plan data while specifically focusing on the relationships between room nodes.
This enables the model to ultimately generate high-quality results that are topologically consistent and reasonable.
Specifically, given the $M(x_t)$, $t$, \& $c$, our GATransformer $\epsilon^{topo}_\theta(M(x_t), t, c)$ can infer the noise required to reconstruct $x_{t-1}$ at time $t-1$:

\begin{equation}
    x_{t-1} = \frac{1}{\sqrt{\alpha_t}}(M(x_t) - \frac{1-\alpha_t}{\sqrt{1-\bar \alpha_t}}\epsilon^{topo}_\theta(M(x_t), t, c))+\sigma_t z
    \label{eq:reverse_c}
\end{equation}

Both $M(x_t)$ and $c$ are independently transformed into $512$-dimensional embeddings using linear layers.
The $t$ is embedded using a sinusoidal function to generate $512$-dimensional positional embeddings that encode temporal information.
These embeddings of $M(x_t)$, $t$, \& $c$ are summed element-wise to create a unified $512$-dimensional representation, and then fed into a stack of four Transformer encoders.
Each encoder comprises instance normalization to stabilize the input features, multi-head attention mechanisms to capture intricate dependencies within the data, and feed-forward networks enhanced with activation functions and dropout for regularization.

Following the Transformer encoders~\cite{Vaswani-17}, the model constructs a fully connected graph for each input sample and processes the encoded features through the GATs~\cite{Velickovic-18-GAT}.
The GATs~\cite{Velickovic-18-GAT} layer applies linear transformations to node features, computes attention scores to assess the importance of neighboring nodes, and aggregates information based on these scores, thereby enriching the feature representations with relational context.
Finally, the enriched features are forwarded through a series of three linear layers interleaved with ReLU activation functions, progressively mapping the features from $512$ to the desired output dimension.

\subsection{Geometry-enhanced mechanism} \label{subsec:method_geometry_module}

The geometric properties of the floor plan are also crucial.
Therefore, we introduce a geometry-enhanced mechanism, which includes three alignment loss functions ($\mathcal{L}_{\mathrm{align}^\mathrm{GT}}$, $\mathcal{L}_{\mathrm{align}^\mathrm{bound}}$, \& $\mathcal{L}_{\mathrm{align}^\mathrm{neigh}}$) that focus on the geometric properties of the generated floor plan to enhance its geometric plausibility.
Among them, $\mathcal{L}_{\mathrm{align}^\mathrm{GT}}$ aims to enable the diffusion model to more accurately capture the data distribution from the input boundary to the floor plan.
The other two alignment losses ($\mathcal{L}_{\mathrm{align}^\mathrm{bound}}$ \& $\mathcal{L}_{\mathrm{align}^\mathrm{neigh}}$) encourage the model to minimize geometrically unreasonable designs, such as overlaps or gaps between rooms and either the input boundary or neighboring rooms.

Specifically, by rewriting Equation~(\ref{eq:forward}), we can obtain $\tilde x_0$, an approximation of $x_0$, at each time step using the noise predicted by ${\epsilon}^{topo}_{\theta}$ as:

\begin{equation}
    \tilde x_0 = (M(x_t) - \sqrt{1-\bar{\alpha}_t} {\epsilon}^{topo}_{\theta}(M(x_t), t, c)) / \sqrt{\bar{ \alpha_t}}
    \label{eq:predicted}
\end{equation}
We first introduce $\mathcal{L}_{\mathrm{align}^\mathrm{GT}}$ to encourage plausible predictions of $x_0$ at each time step $t$ by computing the Mean Squared Error (MSE) between the $\tilde x_0$ and $x_0$:

\begin{equation}
    \mathcal{L}_{\mathrm{align}^\mathrm{GT}} = || \tilde x_{0} (M(x_t), {\epsilon}^{topo}_{\theta}(M(x_t), t, c)) - {x_{0}} ||^2 \cdot \text{Mask}_{\text{cond}}
    \label{eq:align_loss_gt}
\end{equation}

From the predicted $\tilde x_0$, we can extract a set of room box coordinates $r_{i = 1}^n$.
We can also obtain a set of corner coordinates from the input boundary $b_{j = 1}^m$.
We then introduce $\mathcal{L}_{\mathrm{align}^\mathrm{bound}}$ to encourage the model to reduce the gaps between the generated room boxes $r_{i = 1}^n$ and the input boundary $b_{j = 1}^m$:

\begin{equation}
    \mathcal{L}_{\mathrm{align}^\mathrm{bound}} = \mathbb{E}_{j \in m} D (b_j, r_{i = 1}^n)
    \label{eq:align_loss_boundary}
\end{equation}
where $D(\cdot)$ is a distance function, which ensures that if the boundary corner $b_j$ lies within any room, its return value is $0$.
Otherwise, the return value is the distance from $b_j$ to the nearest edge of the closest room box.

The $\mathcal{L}_{\mathrm{align}^\mathrm{neigh}}$ aims to encourage the model to minimize the gaps and overlaps between the generated room boxes $r_{i = 1}^n$:

\begin{equation}
    \mathcal{L}_{\mathrm{align}^\mathrm{neigh}} = \mathbb{E}_{1 \leq i < j \leq n} IoU (r_i, r_j) + Gap (r_i, r_j)
    \label{eq:align_loss_neighbors}
\end{equation}
where the Intersection over Union ($IoU(\cdot)$) is defined as the ratio of the intersection area to the union area between two room boxes ($r_i$ \& $r_j$).
$Gap(\cdot)$ calculates the horizontal and vertical distances between two room boxes ($r_i$ \& $r_j$) and returns the sum of these distances.
If either distance exceeds the set threshold $d$, the function returns $0$.

The overall loss function for training our CE2EPlan is defined as:

\begin{equation}
    \mathcal{L} = \mathcal{L}_{\mathrm{t-1}}
     + \lambda_{1} \mathcal{L}_{\mathrm{align}^\mathrm{GT}}
     + \lambda_{2} \mathcal{L}_{\mathrm{align}^\mathrm{bound}}
     + \lambda_{3} \mathcal{L}_{\mathrm{align}^\mathrm{neigh}}
    \label{eq:total_loss}
\end{equation}
For the boundary-unconstrained floor plan generation, we exclude the $\mathcal{L}_{\mathrm{align}^\mathrm{bound}}$ in training.
At the inference stage, we employ the post-processing steps in~\cite{Hu-20} to further align the generated room boxes, and the rule-based algorithm in~\cite{Wu-19, Hu-20} to add doors and windows for a more refined layout.


\section{Experiments} \label{sec:experiments}

We conduct quantitative comparisons, qualitative evaluations, user studies, and ablation experiments to thoroughly evaluate our method, followed by a detailed discussion.

\begin{table}[t]
    \centering
    \caption{Quantitative comparison between our CE2EPlan and the state-of-the-art methods (RPLAN, ActFloor-GAN, WallPlan, Graph2Plan, $\text{DiffPlanner}_{\text{I}}$, \& $\text{iPLAN}_{\text{I}}$) for automatic floor plan generation only from boundary information ($\text{Mode}_{\text{auto}}$).}
    \label{tab:quant_fp_wboun_auto}
    \resizebox{\columnwidth}{!}{\begin{tabular}{|l|c|cc|}
        \hline
        & Solution & \multicolumn{2}{c|}{Metric}  \\
        Method & Path & FID $(\downarrow)$ & GED $(\downarrow)$  \\
        \hline
        RPLAN~\cite{Wu-19}
        & \begin{tikzpicture}[baseline=(base.base)]
            \node[inner sep=0] (base) at (0,0) {\strut};
            \node (A) at (0,0) {$\alpha$};
            \node (B) at (0.8,0) {$\delta$};
            \node (C) at (1.6,0) {$\lambda$};
            \node (D) at (2.4,0) {$\sigma$};
            \node (E) at (3.2,0) {$\omega$};
            \draw[->] (A) -- (B);
            \draw[->] (B) -- (C);
            \draw[->] (C) -- (D);
            \draw[->] (D) -- (E);
            \draw[->, bend right=20] (D) to (B);
        \end{tikzpicture} & 4.64 & 3.51 \\
        ActFloor-GAN~\cite{Wang-21}
        & \begin{tikzpicture}[baseline=(base.base)]
            \node[inner sep=0] (base) at (0,0) {\strut};
            \node (A) at (0,0) {$\alpha$};
            \node (B) at (0.8,0) {$\theta$};
            \node (C) at (1.6,0) {$\omega$};
            \draw[->] (A) -- (B);
            \draw[->] (B) -- (C);
        \end{tikzpicture} & 4.54 & 3.88 \\
        WallPlan~\cite{Sun-22}
        & \begin{tikzpicture}[baseline=(base.base)]
            \node[inner sep=0] (base) at (0,0) {\strut};
            \node (A) at (0,0) {$\alpha$};
            \node (B) at (0.8,0) {$\gamma$};
            \node (C) at (1.6,0) {$\eta$};
            \node (D) at (2.4,0) {$\rho$};
            \node (E) at (3.2,0) {$\omega$};
            \draw[->] (A) -- (B);
            \draw[->] (B) -- (C);
            \draw[->] (C) -- (D);
            \draw[->] (D) -- (E);
            \draw[->, bend right=20] (E) to (C);
        \end{tikzpicture} & 2.55 & 4.28 \\
        Graph2Plan~\cite{Hu-20}
        & \begin{tikzpicture}[baseline=(base.base)]
				\node[inner sep=0] (base) at (0,0) {\strut};
				\node (A) at (0,0) {$\alpha$};
				\node (B) at (0.8,0) {$\lambda*$};
				\node (C) at (1.6,0) {$\omega$};
				\draw[->] (A) -- (B);
				\draw[->] (B) -- (C);
			\end{tikzpicture} & 2.03 & 3.94 \\
		$\text{DiffPlanner}_{\text{I}}$~\cite{Wang-25}
        & \begin{tikzpicture}[baseline=(base.base)]
		\node[inner sep=0] (base) at (0,0) {\strut};
		\node (A) at (0,0) {$\alpha$};
		\node (B) at (0.8,0) {$\lambda$};
		\node (C) at (1.6,0) {$\lambda*$};
		\node (D) at (2.4,0) {$\omega$};
		\draw[->] (A) -- (B);
		\draw[->] (B) -- (C);
		\draw[->] (C) -- (D);
        \end{tikzpicture} & \textbf{1.08} & 3.54 \\
        $\text{iPLAN}_{\text{I}}$~\cite{He-22}
        & \begin{tikzpicture}[baseline=(base.base)]
				\node[inner sep=0] (base) at (0,0) {\strut};
				\node (A) at (0,0) {$\alpha$};
				\node (B) at (0.8,0) {$\epsilon$};
				\node (C) at (1.6,0) {$\lambda$};
				\node (D) at (2.4,0) {$\omega$};
				\draw[->] (A) -- (B);
				\draw[->] (B) -- (C);
				\draw[->] (C) -- (D);
				\draw[->, bend right=20] (C) to (B);
				\draw[->, bend right=20] (D) to (C);
			\end{tikzpicture} & 5.38 & 4.18 \\
        CE2EPlan
        & \begin{tikzpicture}[baseline=(base.base)]
				\node[inner sep=0] (base) at (0,0) {\strut};
				\node (A) at (0,0) {$\alpha$};
				\node (B) at (0.8,0) {$\omega$};
				\draw[->] (A) -- (B);
        \end{tikzpicture} & \textbf{1.08} & \textbf{3.48} \\
        \hline
    \end{tabular}}
\end{table}

\subsection{Quantitative comparisons} \label{subsec:ex_quantitative}

\paragraph{FID \& GED comparison}
Following~\cite{He-22, Sun-22, Shabani-23, Wang-25}, we use the Fr\'echet Inception Distance (FID) \cite{Heusel-17} score as a global metric to calculate the distribution similarity between real and generated data.
A lower FID $(\downarrow)$ score indicates that the generated data is more similar to the ground truths (GT).
We convert the generated vector floor plans into $512\times512$ raster images with the same colormap for each competing method to calculate the FID score on the test dataset containing $12K$ examples.

To directly compare the topological structure of the generated vector floor plans with the GT, we also compute the Graph Edit Distance (GED)~\cite{sanfeliu1983GED} from the generated data to GT for each sample in the test dataset containing $12K$ examples and then calculate the average.
A lower GED $(\downarrow)$ score indicates a closer topological resemblance to the GT.

Table~\ref{tab:quant_fp_wboun_auto} shows a quantitative comparison of FID and GED scores between our CE2EPlan and the state-of-the-art methods (RPLAN~\cite{Wu-19}, ActFloor-GAN~\cite{Wang-21}, WallPlan~\cite{Sun-22}, Graph2Plan~\cite{Hu-20}, $\text{DiffPlanner}_{\text{I}}$~\cite{Wang-25}, \& $\text{iPLAN}_{\text{I}}$~\cite{He-22}) for automatic floor plan generation only from boundary information ($\text{Mode}_{\text{auto}}$).
Compared to all previous methods, which are based on a multi-step pipeline, our end-to-end approach achieves the lowest FID and GED scores.
This indicates that our CE2EPlan generates results that are globally closer to GT and also exhibit the most similar topological structure.

We further quantitatively compare our method with two typical pipeline-based methods (iPLAN \& DiffPlanner) for boundary-constrained floor plan generation under two user interaction modes, i.e., using room type input ($\text{Mode}_{\text{t}}$) and using both room type \& location input ($\text{Mode}_{\text{t}\&\text{l}}$).
As shown in Table~\ref{tab:quant_fp_wboun_auto} \& Table~\ref{tab:quant_fp_wboun_tl}, compared to iPLAN \& DiffPlanner, which use three separate models to decompose the generation process into three sub-tasks, our approach only needs a single model to support multiple interaction modes, achieving the lowest FID and GED scores and delivering higher-quality results, thereby demonstrating the superiority of our end-to-end generation strategy.

\begin{table}[t]
    \centering
    \caption{Quantitative comparison between our CE2EPlan and the typical pipeline-based methods (iPLAN \& DiffPlanner) for boundary-constrained floor plan generation under two user interaction modes, i.e., using room type input ($\text{Mode}_{\text{t}}$) and using both room type \& location input ($\text{Mode}_{\text{t}\&\text{l}}$).}
    \label{tab:quant_fp_wboun_tl}
    \resizebox{\columnwidth}{!}{\begin{tabular}{|l|cc|cc|}
	\hline
	& Interaction & Solution & \multicolumn{2}{c|}{Metric} \\
	Method & Mode & Path & FID $(\downarrow)$ & GED $(\downarrow)$ \\
	\hline
	$\text{iPLAN}_{\text{II}}$~\cite{He-22} & $\text{Mode}_{\text{t}}$
	& \begin{tikzpicture}[baseline=(base.base)]
			\node[inner sep=0] (base) at (0,0) {\strut};
			\node (B) at (0.0,0) {$\epsilon$};
			\node (C) at (0.8,0) {$\lambda$};
			\node (D) at (1.6,0) {$\omega$};
			\draw[->] (B) -- (C);
			\draw[->] (C) -- (D);
			\draw[->, bend right=20] (C) to (B);
			\draw[->, bend right=20] (D) to (C);
	\end{tikzpicture} & 5.21 & 2.07 \\
	$\text{CE2EPlan}$ & $\text{Mode}_{\text{t}}$
	& \begin{tikzpicture}[baseline=(base.base)]
			\node[inner sep=0] (base) at (0,0) {\strut};
			\node (B) at (0.0,0) {$\epsilon$};
			\node (C) at (0.8,0) {$\omega$};
			\draw[->] (B) -- (C);
	\end{tikzpicture} & \textbf{0.99} & \textbf{1.78} \\
	\hline
	$\text{iPLAN}_{\text{III}}$~\cite{He-22} & $\text{Mode}_{\text{t}\&\text{l}}$
	& \begin{tikzpicture}[baseline=(base.base)]
			\node[inner sep=0] (base) at (0,0) {\strut};
			\node (C) at (0.0,0) {$\lambda$};
			\node (D) at (0.8,0) {$\omega$};
			\draw[->] (C) -- (D);
			\draw[->, bend right=20] (D) to (C);
	\end{tikzpicture} & 2.13 & 0.48 \\
        $\text{DiffPlanner}_{\text{II}}$~\cite{Wang-25} & $\text{Mode}_{\text{t}\&\text{l}}$
	& \begin{tikzpicture}[baseline=(base.base)]
			\node[inner sep=0] (base) at (0,0) {\strut};
		      \node (B) at (0.0,0) {$\lambda$};
		      \node (C) at (0.8,0) {$\lambda*$};
		      \node (D) at (1.6,0) {$\omega$};
		      \draw[->] (B) -- (C);
		      \draw[->] (C) -- (D);
	\end{tikzpicture} & 0.26 & 0.42 \\
	$\text{CE2EPlan}$ & $\text{Mode}_{\text{t}\&\text{l}}$
	& \begin{tikzpicture}[baseline=(base.base)]
			\node[inner sep=0] (base) at (0,0) {\strut};
			\node (B) at (0.0,0) {$\lambda$};
			\node (C) at (0.8,0) {$\omega$};
			\draw[->] (B) -- (C);
	\end{tikzpicture} & \textbf{0.24} & \textbf{0.41} \\
	\hline
    \end{tabular}}
\end{table}

\begin{table}[t]
    \centering
    \caption{Quantitative evaluation of the ability of our CE2EPlan for boundary-constrained floor plan generation to predict the complete target from partial inputs ($\text{Mode}_{\text{part}}$). The varying proportions of target information in the partial inputs: 25\%, 50\%, and 75\%.}
    \label{tab:quant_fp_wboun_part}
    \begin{tabular}{|l|cc|cc|}
        \hline
        & Interaction & Solution & \multicolumn{2}{c|}{Metric}  \\
        Method & Mode &  Path & FID $(\downarrow)$  & GED $(\downarrow)$  \\
        \hline
        \multirow{3}{*}{$\text{CE2EPlan}$} & 25\%~$\text{Mode}_{\text{part}}$ 
        & \begin{tikzpicture}[baseline=(base.base)]
			\node[inner sep=0] (base) at (0,0) {\strut};
			\node (B) at (0.0,0) {$\omega_{25\%}$};
			\node (C) at (0.8,0) {$\omega$};
			\draw[->] (B) -- (C);
		\end{tikzpicture} & 0.72 & 3.00 \\
        & 50\%~$\text{Mode}_{\text{part}}$
        & \begin{tikzpicture}[baseline=(base.base)]
			\node[inner sep=0] (base) at (0,0) {\strut};
			\node (B) at (0.0,0) {$\omega_{50\%}$};
			\node (C) at (0.8,0) {$\omega$};
			\draw[->] (B) -- (C);
		\end{tikzpicture} & 0.44 & 2.21 \\
        & 75\%~$\text{Mode}_{\text{part}}$
        & \begin{tikzpicture}[baseline=(base.base)]
			\node[inner sep=0] (base) at (0,0) {\strut};
			\node (B) at (0.0,0) {$\omega_{75\%}$};
			\node (C) at (0.8,0) {$\omega$};
			\draw[->] (B) -- (C);
		\end{tikzpicture} & \textbf{0.28} & \textbf{1.09} \\
	\hline
    \end{tabular}
\end{table}

\begin{table}[t]
    \centering
    \caption{Quantitative comparison between our CE2EPlan and the state-of-the-art methods (HouseDiffusion \& DiffPlanner) for boundary-unconstrained floor plan generation under four user interaction modes, i.e., without user input ($\text{Mode}_{\text{auto}}$), using room type input ($\text{Mode}_{\text{t}}$), using both room type \& location input ($\text{Mode}_{\text{t}\&\text{l}}$), and partial input ($\text{Mode}_{\text{part}}$). $*$ indicates that additional room adjacencies are required as input.}
    \label{tab:quant_fp_woboun}
    \resizebox{\columnwidth}{!}{\begin{tabular}{|l|cc|cc|}
        \hline
        & Interaction & Solution & \multicolumn{2}{c|}{Metric}  \\
        Method & Mode &  Path & FID $(\downarrow)$  & GED $(\downarrow)$  \\
        \hline
        $\text{DiffPlanner}_{\text{I}}$~\cite{Wang-25} & $\text{Mode}_{\text{auto}}$
        & \begin{tikzpicture}[baseline=(base.base)]
				\node[inner sep=0] (base) at (0,0) {\strut};
                \node (A) at (0,0) {$\alpha$};
                \node (B) at (0.8,0) {$\lambda$};
                \node (C) at (1.6,0) {$\lambda*$};
		        \node (D) at (2.4,0) {$\omega$};
                \draw[->] (A) -- (B);
                \draw[->] (B) -- (C);
                \draw[->] (C) -- (D);
                \end{tikzpicture} & 3.38 & 4.53 \\
        CE2EPlan & $\text{Mode}_{\text{auto}}$
        & \begin{tikzpicture}[baseline=(base.base)]
				\node[inner sep=0] (base) at (0,0) {\strut};
				\node (A) at (0,0) {$\alpha$};
				\node (B) at (0.8,0) {$\omega$};
				\draw[->] (A) -- (B);
                \end{tikzpicture} & \textbf{2.39} & \textbf{4.45} \\
        \hline
        HouseDiffusion~\cite{Shabani-23} & $\text{Mode}_{\text{t}}*$
        & \begin{tikzpicture}[baseline=(base.base)]
				\node[inner sep=0] (base) at (0,0) {\strut};
				\node (A) at (0,0) {$\epsilon*$};
				\node (B) at (0.8,0) {$\omega$};
				\draw[->] (A) -- (B);
			\end{tikzpicture} & 29.56 & \textbf{0.39} \\
        CE2EPlan & $\text{Mode}_{\text{t}}$
        & \begin{tikzpicture}[baseline=(base.base)]
				\node[inner sep=0] (base) at (0,0) {\strut};
				\node (A) at (0,0) {$\epsilon$};
				\node (B) at (0.8,0) {$\omega$};
				\draw[->] (A) -- (B);
			\end{tikzpicture} & \textbf{2.23} & 2.09 \\
        \hline
        $\text{DiffPlanner}_{\text{II}}$~\cite{Wang-25} & $\text{Mode}_{\text{t}\&\text{l}}$
	& \begin{tikzpicture}[baseline=(base.base)]
			\node[inner sep=0] (base) at (0,0) {\strut};
		      \node (B) at (0.0,0) {$\lambda$};
		      \node (C) at (0.8,0) {$\lambda*$};
		      \node (D) at (1.6,0) {$\omega$};
		      \draw[->] (B) -- (C);
		      \draw[->] (C) -- (D);
	\end{tikzpicture} & 2.89 & 0.50 \\
        CE2EPlan & $\text{Mode}_{\text{t}\&\text{l}}$
        & \begin{tikzpicture}[baseline=(base.base)]
				\node[inner sep=0] (base) at (0,0) {\strut};
				\node (A) at (0,0) {$\lambda$};
				\node (B) at (0.8,0) {$\omega$};
				\draw[->] (A) -- (B);
			\end{tikzpicture} & \textbf{2.37} & \textbf{0.39} \\
        \hline
        \multirow{3}{*}{CE2EPlan} & 25\%~$\text{Mode}_{\text{part}}$
        & \begin{tikzpicture}[baseline=(base.base)]
				\node[inner sep=0] (base) at (0,0) {\strut};
				\node (A) at (0,0) {$\omega_{25\%}$};
				\node (B) at (0.8,0) {$\omega$};
				\draw[->] (A) -- (B);
			\end{tikzpicture} & 2.26 & 3.88 \\
        & 50\%~$\text{Mode}_{\text{part}}$
        & \begin{tikzpicture}[baseline=(base.base)]
				\node[inner sep=0] (base) at (0,0) {\strut};
				\node (A) at (0,0) {$\omega_{50\%}$};
				\node (B) at (0.8,0) {$\omega$};
				\draw[->] (A) -- (B);
			\end{tikzpicture} & 1.98 & 2.91 \\
        & 75\%~$\text{Mode}_{\text{part}}$
        & \begin{tikzpicture}[baseline=(base.base)]
				\node[inner sep=0] (base) at (0,0) {\strut};
				\node (A) at (0,0) {$\omega_{75\%}$};
				\node (B) at (0.8,0) {$\omega$};
				\draw[->] (A) -- (B);
			\end{tikzpicture} & \textbf{1.76} & \textbf{1.42} \\
        \hline
    \end{tabular}}
\end{table}

We further evaluate the ability of our method to predict the complete target from the partial inputs ($\text{Mode}_{\text{part}}$), given the boundary as input.
We set the proportion of target information in the partial inputs to three values: $25\%$, $50\%$, and $75\%$.
As shown in Table~\ref{tab:quant_fp_wboun_part}, even with partial inputs containing only 25\% of the target information, our method achieves relatively low FID and GED scores.
Furthermore, as the proportion of target information increases, the FID and GED scores further decrease.
These experimental results demonstrate the ability of our method to predict complete layouts from partial inputs, indicating strong support for the iterative interaction with users.

Table~\ref{tab:quant_fp_woboun} shows a quantitative comparison of FID and GED scores between our CE2EPlan and the existing state-of-the-art methods (HouseDiffusion~\cite{Shabani-23} \& DiffPlanner~\cite{Wang-25}) for boundary-unconstrained floor plan generation.

For the controllability of the model, same as the prior existing methods~\cite{Nauata-20, Nauata-21}, HouseDiffusion also manually defines a solution path with an intermediate representation ($\epsilon*\rightarrow\omega$), and it only supports a single interaction mode ($\text{Mode}_{\text{t}}*$), i.e., converting bubble diagrams that encode room types and adjacencies ($\epsilon*$) into floor plans ($\omega$).
However, the floor plan design process does not always start from bubble diagrams ($\epsilon*$), which already contain high-level design constraints.
Users require an AI design tool that can not only generate high-quality results from scratch fully automatically ($\alpha\rightarrow\omega$) but also allow user-specified input conditions and iterative interaction~\cite{Shneiderman-20}.
DiffPlanner introduces a multi-step pipeline using even more intermediate representations ($\alpha\rightarrow\lambda\rightarrow\lambda*\rightarrow\omega$), enabling floor plan generation from scratch with a certain degree of controllability.
However, it follows a fixed solution path and requires training multiple separate models.

In contrast, our method is fully end-to-end, requiring no intermediate representations.
With only a single training process, it supports generating high-quality results across multiple user interaction modes, either without any user input ($\text{Mode}_{\text{auto}}$), or including room type input ($\text{Mode}_{\text{t}}$), room type \& location input ($\text{Mode}_{\text{t}\&\text{l}}$), as well as partial input ($\text{Mode}_{\text{part}}$).

For the quality of the generated results, our CE2EPlan achieves lower FID scores across four interaction modes, indicating that, compared to HouseDiffusion \& DiffPlanner, our results are closer to GT in the overall distribution.
It is worth noting that HouseDiffusion achieves a relatively low GED score because it uses bubble diagrams as input, which are derived from GT and encode room counts and adjacencies, the primary factors for calculating the GED score.
Even without explicitly using room adjacencies input, our method naturally generates results with room adjacencies highly similar to GT, thereby achieving the same GED score as HouseDiffusion, given room type \& location as input ($\text{Mode}_{\text{t}\&\text{l}}$).
This is because when room types and locations are already specified, room adjacencies are essentially determined.

\paragraph{Statistics comparison}

Inspired by~\cite{Sun-22, Wang-25}, we gather various statistical metrics to assess the detailed geometric and topological quality of the generated floor plans.
These metrics include the total number of rooms ($\text{N}^\text{r}$), the number of rooms directly connected to the living room ($\text{C}^\text{l}$), the ratio of rooms connected to the living room to all non-living rooms ($\text{C}^\text{r}$), the percentage of the public area (living room) ($\text{A}^\text{l}$), the percentage of the private area (bedroom) ($\text{A}^\text{b}$), and the percentage of the functional area (all other rooms, i.e., kitchen, bathroom, balcony, \& storage) ($\text{A}^\text{o}$).
Here, $\text{N}^\text{r}$, $\text{C}^\text{l}$, \& $\text{C}^\text{r}$ concern topology, and $\text{A}^\text{l}$, $\text{A}^\text{b}$, \& $\text{A}^\text{o}$ reflect geometry.
We compute the average of each statistical metric across the test dataset containing $12K$ examples and calculate the ratio of these averages to the corresponding ground truth (GT) values.
A ratio close to $1$ suggests that the geometry and topology of the model-generated results are more similar to the GT.

Table~\ref{tab:quantitative_fp_statistics} shows a statistics comparison between our method and the state-of-the-art methods (RPLAN~\cite{Wu-19}, ActFloor-GAN~\cite{Wang-21}, WallPlan~\cite{Sun-22}, Graph2Plan~\cite{Hu-20}, $\text{DiffPlanner}_{\text{I}}$~\cite{Wang-25}, \& $\text{iPLAN}_{\text{I}}$ ~\cite{He-22}) for floor plan generation only from boundary ($\text{Mode}_{\text{auto}}$).
This comparison aims to evaluate the geometric and topological characteristics of the model-generated results.
Compared to the prior existing methods, our approach achieves the best performance in four out of six metrics ($\text{C}^\text{l}$, $\text{A}^\text{l}$, $\text{A}^\text{b}$, \& $\text{A}^\text{o}$), and its performance on $\text{N}^\text{r}$ \& $\text{C}^\text{r}$ is nearly identical to the best scores obtained by WallPlan \& $\text{DiffPlanner}_{\text{I}}$, respectively.
These results indicate that our end-to-end method can generate outputs that most closely resemble the GT in both geometry and topology, surpassing the current state-of-the-art methods.

\begin{table}[t]
    \centering
    \caption{Comparison of topology-related ($\text{N}^\text{r}$, $\text{C}^\text{l}$, \& $\text{C}^\text{r}$) and geometry-related statistics ($\text{A}^\text{l}$, $\text{A}^\text{b}$, \& $\text{A}^\text{o}$) between our CE2EPlan and the state-of-the-art methods (RPLAN, ActFloor-GAN, WallPlan, Graph2Plan, $\text{DiffPlanner}_{\text{I}}$, \& $\text{iPLAN}_{\text{I}}$) for automatic floor plan generation only from boundary information ($\text{Mode}_{\text{auto}}$). We report the ratio calculated based on GT. A ratio close to $1$ suggests that the geometry and topology of the model-generated results are more similar to GT.}
    \label{tab:quantitative_fp_statistics}
    \resizebox{\columnwidth}{!}{\begin{tabular}{|l|ccc|ccc|}
        \hline
        & \multicolumn{3}{c|}{Topology-related} & \multicolumn{3}{c|}{Geometry-related} \\
        Method & $\text{N}^\text{r}_\text{avg}$  & $\text{C}^\text{l}_\text{avg}$  & $\text{C}^\text{r}_\text{avg}$ & $\text{A}^\text{l}_\text{avg}$ & $\text{A}^\text{b}_\text{avg}$ & $\text{A}^\text{o}_\text{avg}$  \\
        \hline
        RPLAN~\cite{Wu-19}		 & 0.869 & 0.851 & 0.997 & 1.045 & 0.922 & 1.023 \\
        ActFloor-GAN~\cite{Wang-21} & 0.904  & 0.864 & 0.965 & 0.902 & 0.985 & 1.214 \\
        WallPlan~\cite{Sun-22}     & \textbf{0.998}  & 0.968 & 0.973 & 0.911 & 1.129 & 1.428 \\
        Graph2Plan~~\cite{Hu-20}   & 0.980  & 0.971 & 0.988 & 1.029 & 1.013 & 0.914 \\
        $\text{DiffPlanner}_{\text{I}}$~\cite{Wang-25}
        & 0.996 & 0.997 & \textbf{1.001} & 0.971 & 1.019 & 1.020 \\
        $\text{iPLAN}_{\text{I}}$~\cite{He-22}
        & 0.938  & 0.980 & 1.049 & 1.131 & 0.899 & 0.856 \\
        CE2EPlan
        & 0.995 & \textbf{0.998} & 1.003 & \textbf{0.988} & \textbf{1.007} & \textbf{1.008} \\
        \hline
    \end{tabular}}
\end{table}

\begin{figure}[t]
    \centering
    \includegraphics[width=\linewidth]{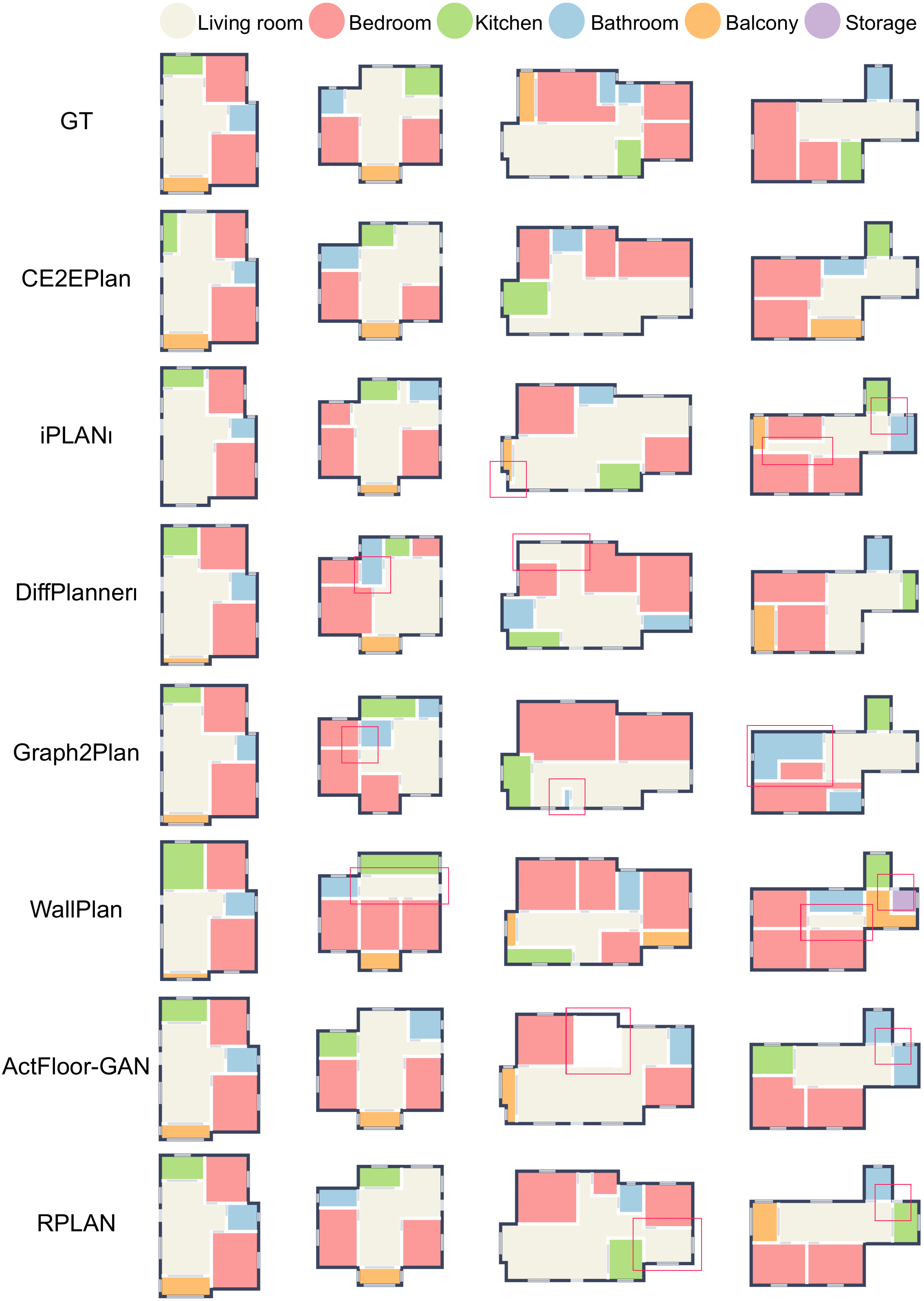}
    \caption{Qualitative comparison on the quality of the generated results between our CE2EPlan and the state-of-the-art methods ($\text{iPLAN}_{\text{I}}$, $\text{DiffPlanner}_{\text{I}}$, Graph2Plan, WallPlan, ActFloor-GAN, \& RPLAN) for floor plan generation only from boundary ($\text{Mode}_{\text{auto}}$). The flawed design is highlighted in the red box.}
    \label{fig:qualitative_previous}
\end{figure}

\subsection{Qualitative evaluations} \label{subsec:ex_qualitative}

\paragraph{Quality}
Figure~\ref{fig:qualitative_previous} presents the qualitative comparison between the generated results of our CE2EPlan and those of the state-of-the-art methods ($\text{iPLAN}_{\text{I}}$~\cite{He-22}, $\text{DiffPlanner}_{\text{I}}$~\cite{Wang-25}, Graph2Plan~\cite{Hu-20}, WallPlan~\cite{Sun-22}, ActFloor-GAN~\cite{Wang-21}, \& RPLAN~\cite{Wu-19}) for automatic floor plan generation from boundary only ($\text{Mode}_{\text{auto}}$).
The first column displays some reasonable results generated by the existing methods, while the subsequent columns show some less realistic generated results.

RPLAN relies on a manually defined iterative prediction process: it first locates the living room, then sequentially determines the other rooms, followed by defining the walls within the boundary ($\begin{tikzpicture}[baseline=(base.base)]
				\node[inner sep=0] (base) at (0,0) {\strut};
				\node (A) at (0,0) {$\alpha$};
				\node (B) at (0.8,0) {$\delta$};
				\node (C) at (1.6,0) {$\lambda$};
				\node (D) at (2.4,0) {$\sigma$};
				\node (E) at (3.2,0) {$\omega$};
				\draw[->] (A) -- (B);
				\draw[->] (B) -- (C);
				\draw[->] (C) -- (D);
				\draw[->] (D) -- (E);
				\draw[->, bend right=20] (D) to (B);
			\end{tikzpicture}$).
Each step impacts subsequent predictions, leading to some issues, such as unutilized corners, incorrect walls, and mislocated rooms.
ActFloor-GAN predicts the human-activity maps within the boundary and then converts them into the floor plans ($\begin{tikzpicture}[baseline=(base.base)]
				\node[inner sep=0] (base) at (0,0) {\strut};
				\node (A) at (0,0) {$\alpha$};
				\node (B) at (0.8,0) {$\theta$};
				\node (C) at (1.6,0) {$\omega$};
				\draw[->] (A) -- (B);
				\draw[->] (B) -- (C);
			\end{tikzpicture}$).
If the model outputs unreasonable human-activity maps, the resulting floor plans may also exhibit issues, such as the unutilized area and the room that obstructs the entrance.
WallPlan generates floor plans by incrementally predicting wall corners ($\begin{tikzpicture}[baseline=(base.base)]
				\node[inner sep=0] (base) at (0,0) {\strut};
				\node (A) at (0,0) {$\alpha$};
				\node (B) at (0.8,0) {$\gamma$};
				\node (C) at (1.6,0) {$\eta$};
				\node (D) at (2.4,0) {$\rho$};
				\node (E) at (3.2,0) {$\omega$};
				\draw[->] (A) -- (B);
				\draw[->] (B) -- (C);
				\draw[->] (C) -- (D);
				\draw[->] (D) -- (E);
				\draw[->, bend right=20] (E) to (C);
			\end{tikzpicture}$), often resulting in structurally complex layouts and tightly packed living rooms.

Graph2Plan generates floor plans from the bubble diagrams encoding the types, locations, \& adjacencies of rooms ($\begin{tikzpicture}[baseline=(base.base)]
				\node[inner sep=0] (base) at (0,0) {\strut};
				\node (A) at (0,0) {$\alpha$};
				\node (B) at (0.8,0) {$\lambda*$};
				\node (C) at (1.6,0) {$\omega$};
				\draw[->] (A) -- (B);
				\draw[->] (B) -- (C);
			\end{tikzpicture}$).
However, these bubble diagrams are retrieved from ground truths (GT) data based on the similarity of the input boundary, and may not always be suitable, leading to some unrealistic designs.
$\text{DiffPlanner}_{\text{I}}$ predicts room nodes and their adjacencies from the input boundary to form the bubble diagrams, and then performs room partitioning to generate the final floor plan ($\begin{tikzpicture}[baseline=(base.base)]
				\node[inner sep=0] (base) at (0,0) {\strut};
                \node (A) at (0,0) {$\alpha$};
                \node (B) at (0.8,0) {$\lambda$};
                \node (C) at (1.6,0) {$\lambda*$};
		        \node (D) at (2.4,0) {$\omega$};
                \draw[->] (A) -- (B);
                \draw[->] (B) -- (C);
                \draw[->] (C) -- (D);
                \end{tikzpicture}$).
However, the multi-module pipeline can introduce issues, such as bedrooms potentially not connected to the living room and underutilized corner spaces.
$\text{iPLAN}_{\text{I}}$ similarly decomposes the generation process, supporting iterative design to correct unreasonable outputs from the preceding stages ($\begin{tikzpicture}[baseline=(base.base)]
				\node[inner sep=0] (base) at (0,0) {\strut};
				\node (A) at (0,0) {$\alpha$};
				\node (B) at (0.8,0) {$\epsilon$};
				\node (C) at (1.6,0) {$\lambda$};
				\node (D) at (2.4,0) {$\omega$};
				\draw[->] (A) -- (B);
				\draw[->] (B) -- (C);
				\draw[->] (C) -- (D);
				\draw[->, bend right=20] (C) to (B);
				\draw[->, bend right=20] (D) to (C);
			\end{tikzpicture}$).
However, it occasionally produces some unrealistic designs in fully automated mode.

Overall, the multi-step pipeline with the fixed solution path can yield reasonable designs, especially for simple common cases in the dataset (e.g., column $1$).
However, these methods are constrained by predefined paths from input to output.
When faced with data that does not nicely align with their pre-established prediction processes, these methods often produce less optimal designs (e.g., column $2-4$).
Our end-to-end learning approach effectively addresses these limitations by allowing the model to learn how to generate floor plans directly from input boundaries without relying on a fixed prediction process.
The qualitative comparison demonstrates that our end-to-end method is more robust than previous methods, capable of producing higher-quality and more realistic results.

\begin{figure}[t]
    \centering
    \includegraphics[width=\linewidth]{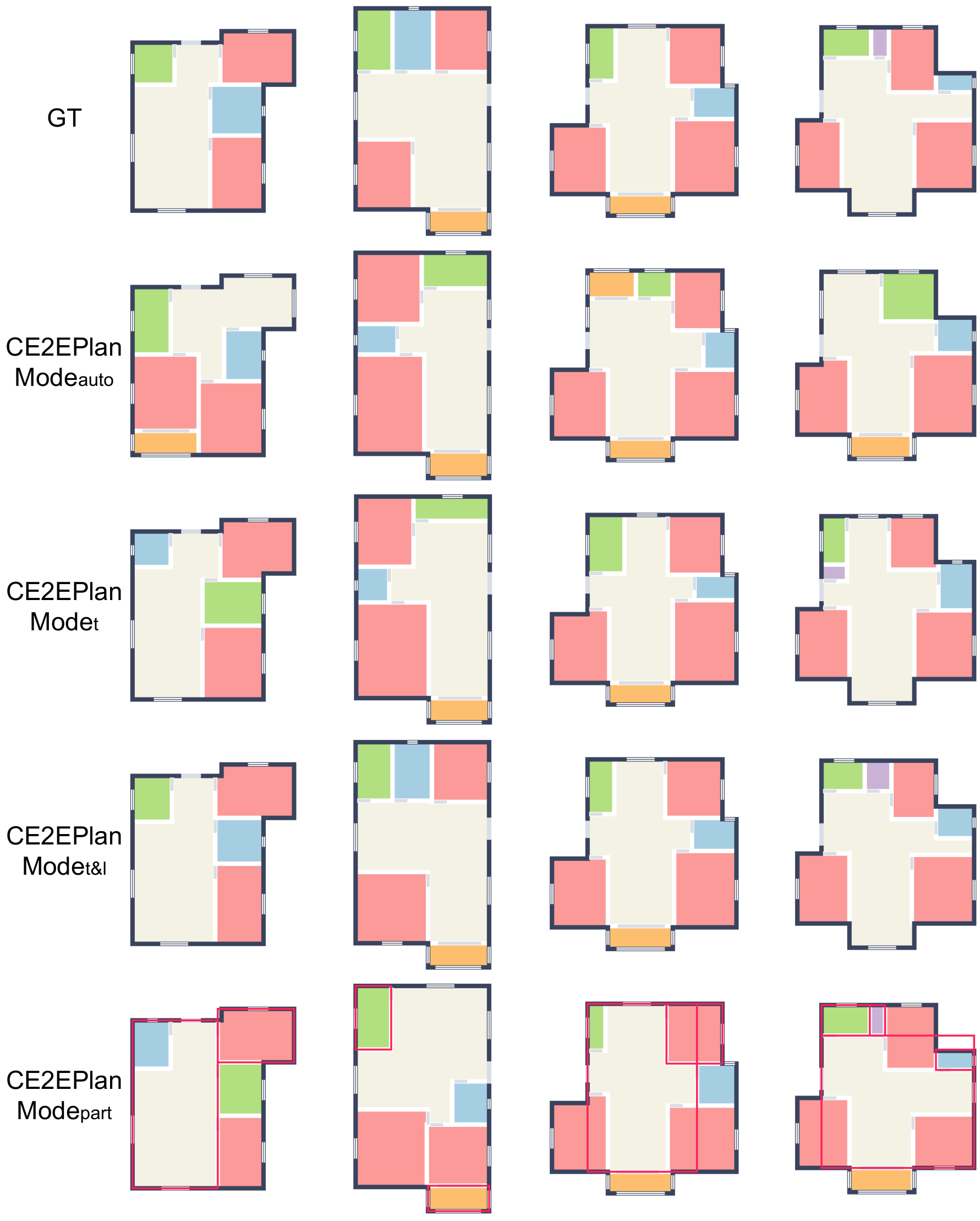}
    \caption{The controllability of our CE2EPlan for the same input boundary across four user interaction modes, i.e., without user input ($\text{Mode}_{\text{auto}}$), using room type input ($\text{Mode}_{\text{t}}$), using both room type \& location input ($\text{Mode}_{\text{t}\&\text{l}}$), and partial input (highlighted with red boxes, $\text{Mode}_{\text{part}}$).}
    \label{fig:qualitative_control}
\end{figure}

\begin{figure}[t]
    \centering
    \includegraphics[width=\linewidth]{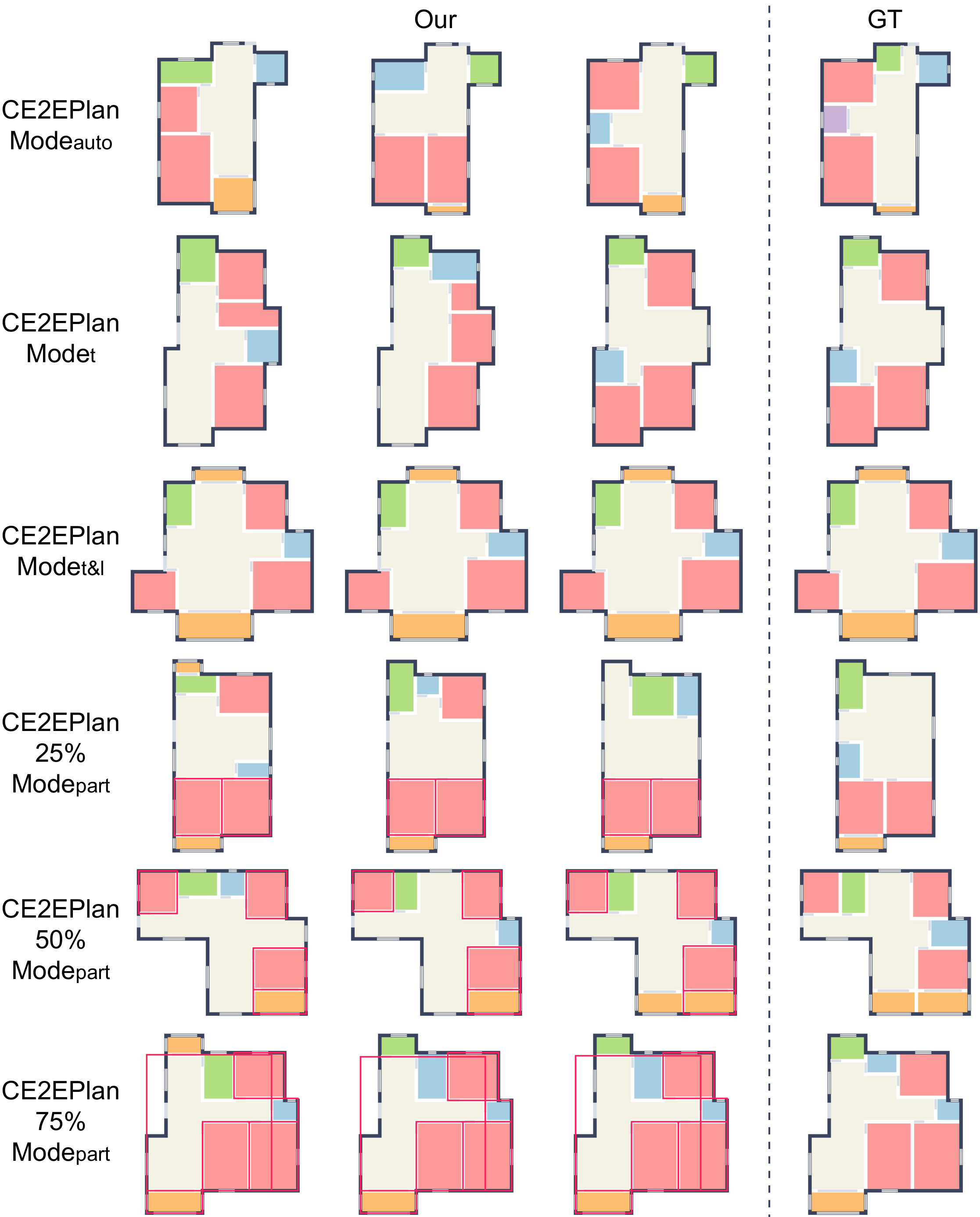}
    \caption{The output diversity of our CE2EPlan for a single input boundary across various user interaction modes, i.e., without user input ($\text{Mode}_{\text{auto}}$), using room type input ($\text{Mode}_{\text{t}}$), using both room type \& location input ($\text{Mode}_{\text{t}\&\text{l}}$), and partial input (highlighted with red boxes) with varying proportions of target information ($25\%~\text{Mode}_{\text{part}}$, $50\%~\text{Mode}_{\text{part}}$, \& $75\%~\text{Mode}_{\text{part}}$).}
    \label{fig:qualitative_diversity}
\end{figure}

\paragraph{Controllability}
Figure~\ref{fig:qualitative_control} presents several groups of floor plans generated by our CE2EPlan based on the same input boundary but with varying user inputs.
Our method performs exceptionally well across all four user interaction modes, i.e., without user input ($\text{Mode}_{\text{auto}}$), using room type input ($\text{Mode}_{\text{t}}$), using room type \& location input ($\text{Mode}_{\text{t}\&\text{l}}$), as well as partial input (highlighted with red boxes in Figure~\ref{fig:qualitative_control}, $\text{Mode}_{\text{part}}$).
In each mode, our method consistently generates high-quality floor plans that meet user specifications while remaining realistic and coherent.
Notably, unlike prior methods that use a multi-step pipeline, our CE2EPlan is end-to-end, requiring only a single training phase while still achieving a rich degree of controllability, demonstrating flexibility and robustness.

Moreover, when handling the two very similar boundaries (e.g., column $3$ \& $4$), our method accurately identifies subtle differences in the boundaries and adapts the output layout accordingly, even though we do not use a dedicated image-based network to extract features from the rasterized boundary as others do~\cite{Wu-19, Hu-20, Wang-21, He-22, Sun-22}.
Instead, we feed the vector boundary directly into the model, demonstrating its capability to discern fine boundary details and generate layouts that align precisely with the input boundary.

\paragraph{Diversity}
The existing methods achieve diversity in generated results by altering the output of the preceding model to affect subsequent ones, relying on the sequential prediction process.
For example, they vary the order of generated room types~\cite{He-22}, adjust the iterative prediction of room locations~\cite{Wu-19}, or modify intermediate inputs like human-activity maps~\cite{Wang-21}, wall masks~\cite{Sun-22}, and bubble diagrams~\cite{Hu-20, Wang-25}.
In contrast, our approach achieves genuine diversity.
Since the reverse process of our CE2EPlan starts from a standard Gaussian random variable $x_T\sim\mathcal{N}(0,\mathbf{I})$, it can generate diverse results from the same input, offering users a range of options.
Additionally, our method supports diverse outputs under various user inputs, as shown in Figure~\ref{fig:qualitative_diversity}.

Note that the diversity decreases as more user input information is provided, which is expected since additional conditions reduce the design space.
In two modes ($\text{Mode}_{\text{auto}}$ \& $\text{Mode}_{\text{t}}$), the model can generate diverse results.
However, when room locations are also specified, the model can only produce results with slight size variations ($\text{Mode}_{\text{t}\&\text{l}}$).
It is understandable that even human designers have limited flexibility once room types and locations are fixed, leaving room size as the primary adjustable factor.
Similarly, as the fixed rooms with red boxes in Figure~\ref{fig:qualitative_diversity} increase under $\text{Mode}_{\text{part}}$, the space for the model-controlled variation decreases, e.g., when two adjacent bedrooms are specified in the $25\%~\text{Mode}_{\text{part}}$, the empty area below them would most logically be designed as a balcony for a coherent layout.
Consequently, the model utilizes the upper area to introduce output diversity.

\begin{figure}[t]
    \centering
    \includegraphics[width=\linewidth]{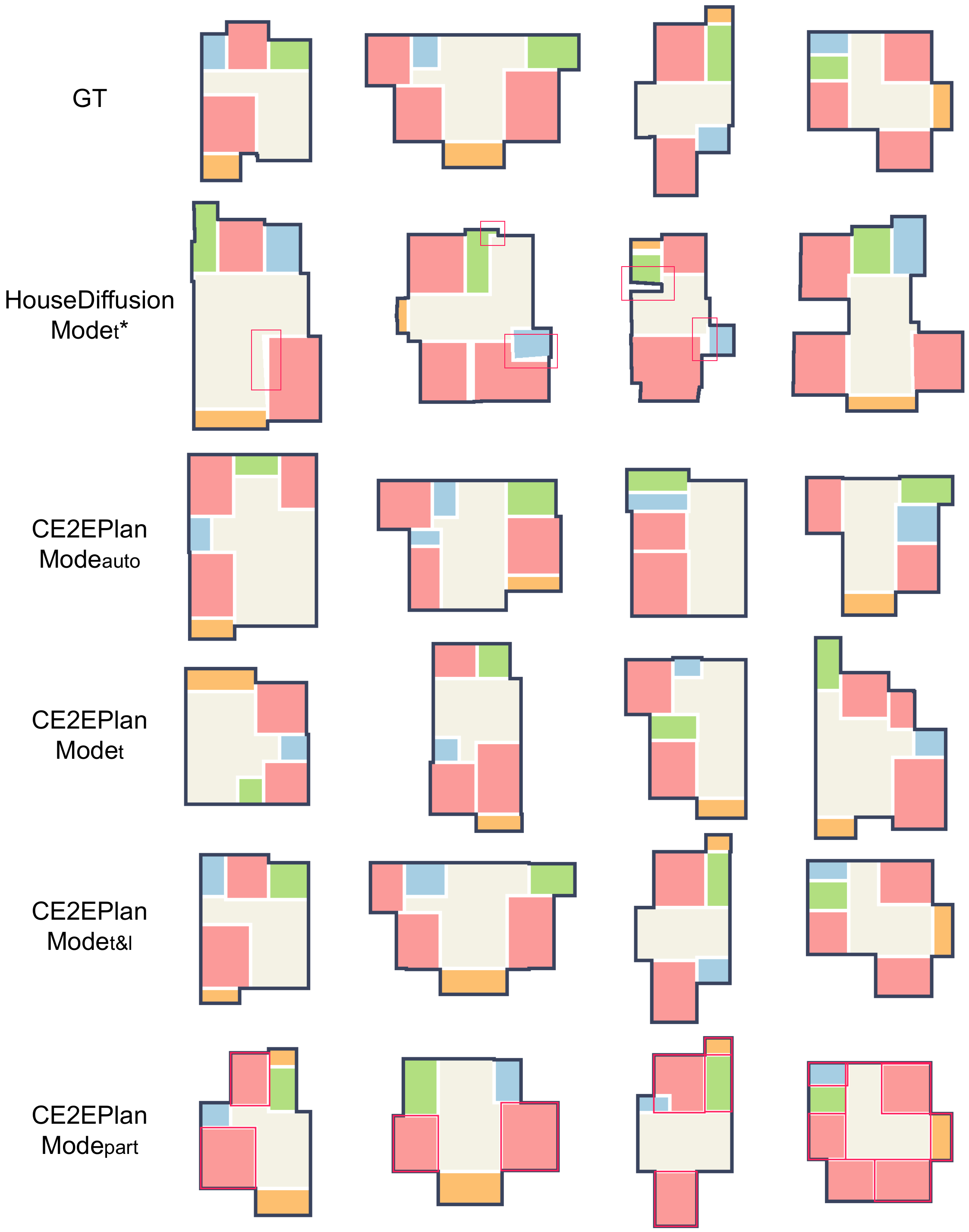}
    \caption{Qualitative comparison between our CE2EPlan and HouseDiffusion for boundary-unconstrained floor plan generation. Our CE2EPlan supports four user interaction modes, i.e., without user input ($\text{Mode}_{\text{auto}}$), using room type input ($\text{Mode}_{\text{t}}$), using both room type \& location input ($\text{Mode}_{\text{t}\&\text{l}}$), and partial input ($\text{Mode}_{\text{part}}$), whereas HouseDiffusion only supports a single interaction mode ($\text{Mode}_{\text{t}}*$). $*$ indicates that additional room adjacencies are required as input.}
    \label{fig:qualitative_control_woboun}
\end{figure}

\begin{figure}[t]
    \centering
    \includegraphics[width=\linewidth]{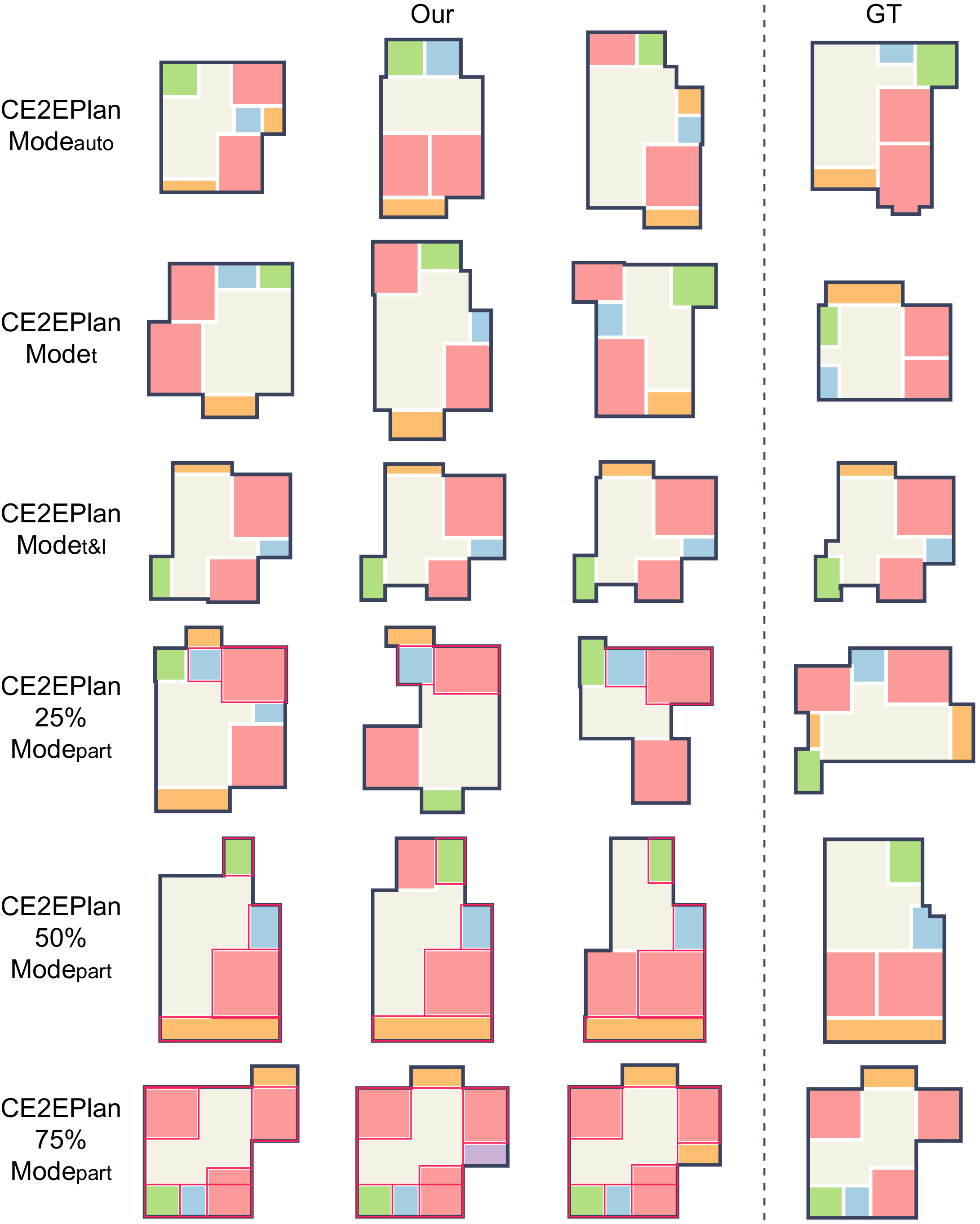}
    \caption{The output diversity of our CE2EPlan in boundary-unconstrained floor plan generation for a single input across various user interaction modes, i.e., without user input ($\text{Mode}_{\text{auto}}$), input room type ($\text{Mode}_{\text{t}}$), input room type \& location ($\text{Mode}_{\text{t}\&\text{l}}$), and partial input (highlighted with red boxes) with varying proportions of target information ($25\%~\text{Mode}_{\text{part}}$, $50\%~\text{Mode}_{\text{part}}$, \& $75\%~\text{Mode}_{\text{part}}$).}
    \label{fig:qualitative_diversity_woboun}
\end{figure}

\paragraph{Comparison with HouseDiffusion}
Figure~\ref{fig:qualitative_control_woboun} presents the qualitative comparison for boundary-unconstrained floor plan generation between our CE2EPlan and HouseDiffusion~\cite{Shabani-23}.
Note that the exterior wall boundaries shown in the floor plans are not provided as inputs to the model.
They are extracted from the final layouts generated by the model solely for visual consistency.
Since HouseDiffusion uses bubble diagrams that encode room types and adjacencies from GT as input ($\text{Mode}_{\text{t}}*$), its generated results achieve topological consistency with the GT.
However, it often produces geometrically unreasonable designs, such as distorted room shapes, irregular room edges, and large gaps between rooms.
In contrast, our method generates results that are both geometrically and topologically reasonable across all four user interaction modes, i.e., without user input ($\text{Mode}_{\text{auto}}$), input room type ($\text{Mode}_{\text{t}}$), input room type \& location ($\text{Mode}_{\text{t}\&\text{l}}$), and partial input (highlighted with red boxes in Figure~\ref{fig:qualitative_control_woboun}, $\text{Mode}_{\text{part}}$).

This discrepancy may arise from the fact that HouseDiffusion represents rooms as a set of corner points, significantly increasing the burden on the model to fit the data distribution.
In contrast, we follow prior concepts~\cite{He-22, Hu-20, Wang-25} by using the bounding boxes to represent the rooms to simplify the prediction targets as much as possible.
This allows our model to more effectively capture the data distribution.
Additionally, our multi-condition masking mechanism supports easy user input, the topology-enhanced module ensures the model captures the topological features of floor plans, and the geometry-enhanced module as well as the post-processing steps improve the geometric plausibility of the generated results.

Figure~\ref{fig:qualitative_diversity_woboun} demonstrates that our CE2EPlan can generate multiple results for a single input based on various user inputs, i.e., without user input ($\text{Mode}_{\text{auto}}$), input room type ($\text{Mode}_{\text{t}}$), input room type \& location ($\text{Mode}_{\text{t}\&\text{l}}$), and partial input (highlighted with red boxes) with varying proportions of target information ($25\%~\text{Mode}_{\text{part}}$, $50\%~\text{Mode}_{\text{part}}$, \& $75\%~\text{Mode}_{\text{part}}$).
In comparison, although HouseDiffusion can also produce multiple outputs for the same input, it only supports a single type of user input, i.e., the bubble diagrams ($\text{Mode}_{\text{t}}*$).
Furthermore, compared to the performance of boundary-constrained floor plan generation, the absence of boundary constraints provides the model with more design space to achieve more output diversity.
However, the output diversity still decreases as the amount of user input information increases.

\begin{figure}[t]
	\centering
	\includegraphics[width=\linewidth]{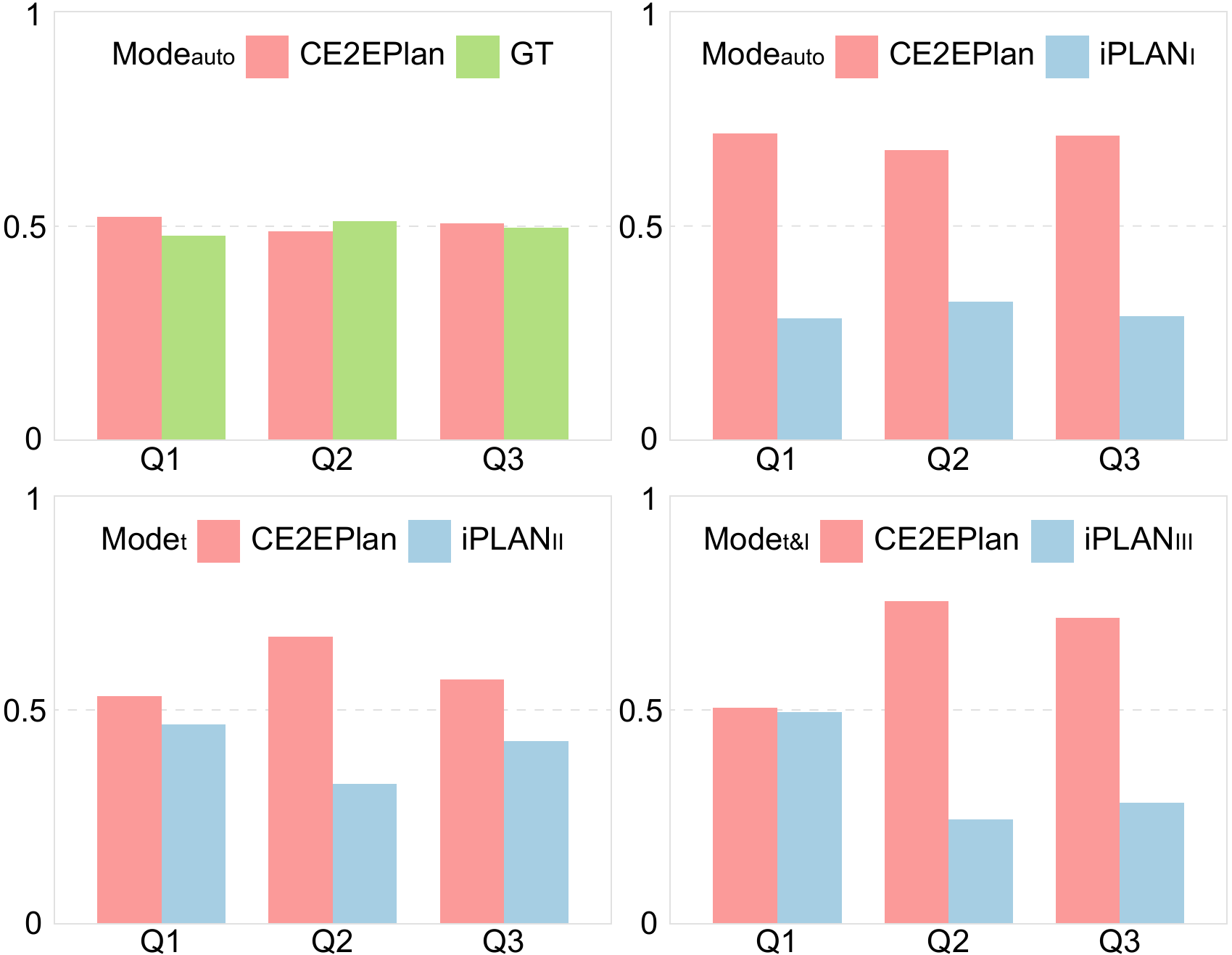}
	\caption{The results of the perceptual studies indicate that the floor plans generated by our CE2EPlan are comparable to the ground truths (GT) and outperform the results from the pipeline-based method (iPLAN).}
	\label{fig:userstudy}
\end{figure}

\subsection{User studies} \label{subsec:ex_userstudy}

We conduct perceptual studies with designers to assess whether the results generated by our CE2EPlan without user input ($\text{Mode}_{\text{auto}}$) are comparable to the real-world designs from ground truths (GT), and whether they outperform the typical pipeline-based method iPLAN~\cite{He-22} under different user interaction modes, i.e., without user input ($\text{Mode}_{\text{auto}}$), input room type ($\text{Mode}_{\text{t}}$), and input room type \& location ($\text{Mode}_{\text{t}\&\text{l}}$), for the boundary-constrained generation task, as shown in Figure~\ref{fig:userstudy}.

We set four groups of experiments, i.e., CE2EPlan under $\text{Mode}_{\text{auto}}$ vs. GT, CE2EPlan under $\text{Mode}_{\text{auto}}$ vs. $\text{iPLAN}_{\text{I}}$ under $\text{Mode}_{\text{auto}}$, CE2EPlan under $\text{Mode}_{\text{t}}$ vs. $\text{iPLAN}_{\text{II}}$ under $\text{Mode}_{\text{t}}$, \& CE2EPlan under $\text{Mode}_{\text{t}\&\text{l}}$ vs. $\text{iPLAN}_{\text{III}}$ under $\text{Mode}_{\text{t}\&\text{l}}$.
Each group includes $30$ comparison tasks, where each pair of floor plans consists of one from the GT or iPLAN and one generated by our CE2EPlan.

Three designers with extensive experience in floor plan design are invited to participate in the perceptual studies.
We design three questions focusing on different aspects to comprehensively evaluate the floor plans, i.e., (Q1) Which one is better designed in terms of topology, primarily determined by the number of rooms and their connections?
(Q2) Which one is better designed in terms of geometry, primarily determined by room sizes and shapes?
(Q3) Overall, which one is better designed?

The participants are asked to complete a forced-choice comparison task by rating each pair as ``better/worse/equally good'' for questions Q1$\sim$Q3, and are not aware of the origin of the floor plans (being from CE2EPlan, GT, or iPLAN).
In total, we obtain $90$ answers ($30$ pairs $\times$ $3$ participants) for each group, and use the winning percentage to analyze the voting results.
This is defined as the number of wins divided by the total number of forced-choice comparison tasks completed.
A win (``better'') counts as a score of $1$, a loss (``worse'') counts as a score of $0$, and a draw (``equally good'') counts as $1/2$.

Figure~\ref{fig:userstudy} presents the results of the perceptual studies.
Compared to GT, the results generated by our CE2EPlan are comparable in terms of topology, geometry, and overall design quality.
Compared to the pipeline-based method $\text{iPLAN}_{\text{I}}$, our CE2EPlan outperforms it in topology, geometry, and overall design quality under $\text{Mode}_{\text{auto}}$.
Under $\text{Mode}_{\text{t}}$, our CE2EPlan shows a slight advantage over $\text{iPLAN}_{\text{II}}$ in topology, while still achieving better results in geometry and overall design quality.
Under $\text{Mode}_{\text{t}\&\text{l}}$, our CE2EPlan performs on par with $\text{iPLAN}_{\text{III}}$ in topology but clearly excels in both geometry and overall design quality.
This is understandable, as the design space for topology becomes increasingly constrained when room types and room locations are progressively provided as user input.

\begin{table*}[t]
    \centering
    \caption{Ablation study on the impact of the multi-condition masking (Mask), topology-enhanced (Topo), geometry-enhanced (Geom), and post-processing (Post) modules on the overall model (CE2EPlan) using the topology-related, geometry-related, and alignment-related metrics. For topology- and geometry-related metrics, values approaching $1$ suggest stronger model performance. For alignment-related metrics, lower values indicate better model performance.}
    \label{tab:ablation_fp}
    \resizebox{\linewidth}{!}{\begin{tabular}{|l|cccc|ccc|ccc|cc|}
        \hline
        &\multicolumn{4}{c|}{Module}&\multicolumn{3}{c|}{Topology-related} & \multicolumn{3}{c|}{Geometry-related} &\multicolumn{2}{c|}{Alignment-related} \\
			& w/ Mask & w/ Topo & w/ Geom & w/ Post & $\text{N}^\text{r}_\text{avg}$  & $\text{C}^\text{l}_\text{avg}$  & $\text{C}^\text{r}_\text{avg}$ & $\text{A}^\text{l}_\text{avg}$ & $\text{A}^\text{b}_\text{avg}$ & $\text{A}^\text{o}_\text{avg}$ & $\mathcal{L}_\mathrm{align}{^\mathrm{bound}_\text{avg}}(\downarrow)$ & $\mathcal{L}_\mathrm{align}{^\mathrm{neigh}_\text{avg}}(\downarrow)$   \\
        \hline
        $\text{CE2EPlan}$ &\cmark  & \cmark & \cmark  & \cmark & \textbf{0.995} & \textbf{0.998} & \textbf{1.003} & \textbf{0.988} & \textbf{1.007} & \textbf{1.008} & \textbf{0.0003} & \textbf{0.0062} \\
        $\text{CE2EPlan}_{\text{I}}$&\cmark  & \cmark & \cmark  & \xmark & \textbf{0.995} & 0.994 & 0.996 & 1.060 & 0.914 & 0.857 & 0.0039 & 0.0103 \\
        $\text{CE2EPlan}_{\text{II}}$&\cmark  & \cmark & \xmark  & \xmark & \textbf{0.995} & 0.992 & \textbf{0.997} & 1.086 & 0.889 & 0.822 & 0.0051 & 0.0166 \\
        $\text{CE2EPlan}_{\text{III}}$&\cmark  & \xmark & \xmark  & \xmark & 0.984 & 0.975 & 0.981 & 1.091 & 0.890 & 0.791 & 0.0058 & 0.0163 \\
        $\text{CE2EPlan}_{\text{IV}}$&\xmark  & \xmark & \xmark  & \xmark & 1.006 & 0.988 & 0.969 & 1.111 & 0.874 & 0.781 & 0.0060 & 0.0172 \\
        \hline
    \end{tabular}}
\end{table*}

We further design a lightweight interactive experiment to more directly evaluate the controllability and practical usability of CE2EPlan in real design workflows.
For each trial, a random test sample is presented to a participant, who is asked to simulate an early-stage design session by specifying constraints incrementally.
The user first selects whether the generation should be boundary-free or boundary-constrained.
In the latter case, the ground-truth boundary is provided as the input condition.
The participant then chooses one of the four interaction modes supported by CE2EPlan.
\begin{itemize}
	\item $\text{Mode}_{\text{auto}}$: The system produces layouts without additional user input.
	\item $\text{Mode}_{\text{t}}$: The user specifies the desired room types and their counts.
	\item $\text{Mode}_{\text{t}\&\text{l}}$: The user additionally provides approximate room locations.
	\item $\text{Mode}_{\text{part}}$: The user selects a subset of rooms from the ground-truth layout to remain fixed, and the model completes the rest.
\end{itemize}
Given the specified constraints, CE2EPlan generates five candidate layouts for the user to inspect.

The participant then evaluates our model using a structured questionnaire based on a five-point Likert scale (1 = strongly disagree, 2 = disagree, 3 = neutral, 4 = agree, 5 = strongly agree).
The questionnaire covers three aspects:
\begin{itemize}
	\item Controllability: ``The provided interaction modes are meaningful and useful in realistic design scenarios.''
	\item Generation quality: ``The produced layouts are high-quality and reasonable under the given constraints.''
	\item Diversity: ``The model offers sufficient variation to support genuine design exploration.''
\end{itemize}

For each aspect, we collect $30$ responses ($10$ trials $\times$ $3$ participants).
The interactive study yields highly positive feedback, with average ratings of $4.50$ for controllability, $4.73$ for generation quality, and $4.20$ for diversity and creativity.
These results clearly demonstrate that CE2EPlan not only produces high-quality layouts but also provides intuitive and effective controllability to support interactive design workflows.

\subsection{Ablation experiments} \label{subsec:ex_ablation}

We use topology-related, geometry-related, and alignment-related metrics to evaluate the impact of the multi-condition masking (Mask), topology-enhanced (Topo), geometry-enhanced (Geom), and post-processing (Post) modules on the overall model (CE2EPlan) for floor plan generation only from boundary input ($\text{Mode}_{\text{auto}}$), as shown in Table~\ref{tab:ablation_fp}.
The calculation of the topology-related ($\text{N}^\text{r}$, $\text{C}^\text{l}$, \& $\text{C}^\text{r}$) and geometry-related metrics ($\text{A}^\text{l}$, $\text{A}^\text{b}$, \& $\text{A}^\text{o}$) is described in Section~\ref{subsec:ex_quantitative}.
A value close to $1$ indicates that the geometry and topology of the model-generated results are more similar to the GT.
We also compute the average of the two alignment losses ($\mathcal{L}_{\mathrm{align}^\mathrm{bound}}$ \& $\mathcal{L}_{\mathrm{align}^\mathrm{neigh}}$) introduced in Section~\ref{subsec:method_geometry_module} across the test dataset containing $12K$ examples as alignment-related metrics, where lower values indicate lower alignment losses and better performance.

When the post-processing module (Post) is removed, we observe a decrease in performance on the geometry-related and alignment-related metrics, while the topology-related metrics show little change ($\text{CE2EPlan}_{\text{I}}$ vs. $\text{CE2EPlan}$).
Further removing the geometry-enhanced module (Geom) results in minimal change in the topology-related metrics but a further drop in the geometry-related and alignment-related metrics ($\text{CE2EPlan}_{\text{II}}$ vs. $\text{CE2EPlan}_{\text{I}}$).
This indicates that the Geom and Post effectively improve the geometric coherence of the generated results, successfully addressing misalignments between room boxes.

When we further remove the topology-enhanced module (Topo), disabling the use of Graph Attention Networks (GATs)~\cite{Velickovic-18-GAT} for capturing room-to-room topological relationships, the performance of $\text{CE2EPlan}_{\text{III}}$ on the topology-related metrics declines significantly, while the geometry-related and alignment-related metrics remain similar to those in $\text{CE2EPlan}_{\text{II}}$ ($\text{CE2EPlan}_{\text{III}}$ vs. $\text{CE2EPlan}_{\text{II}}$).
Finally, we conduct an ablation of the multi-condition masking module (Mask).
Compared with $\text{CE2EPlan}_{\text{III}}$ including the masking step, the performance of $\text{CE2EPlan}_{\text{IV}}$ across various metrics is roughly comparable ($\text{CE2EPlan}_{\text{IV}}$ vs. $\text{CE2EPlan}_{\text{III}}$).
This suggests that our masking successfully adds controllability to the model without diminishing its generative capacity.

\begin{figure}[t]
	\centering
	\includegraphics[width=\linewidth]{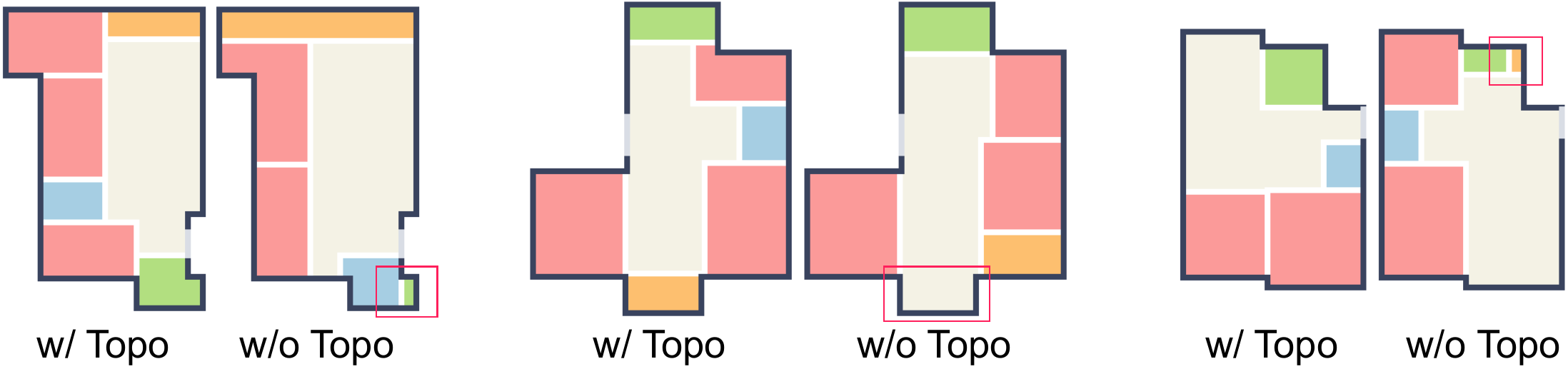}
	\caption{Qualitative comparison between our full model CE2EPlan (w/ Topo) and a variant without the topology-enhanced module (w/o Topo). The flawed design is highlighted in the red box.}
	\label{fig:ex_quali_ablation}
\end{figure}

To further evaluate the effectiveness of the GATs, we conduct a qualitative comparison between our full model CE2EPlan (w/ Topo) and a variant without the topology-enhanced module (w/o Topo), thus generating floor plans only from boundary information ($\text{Mode}_{\text{auto}}$).
As shown in Figure~\ref{fig:ex_quali_ablation}, removing the GATs leads the model (w/o Topo) to generate incorrect adjacency structures, resulting in semantically unreasonable layouts.
In the left example, the kitchen is incorrectly placed adjacent to the bathroom and squeezed into the lower-right corner, producing an unrealistically small kitchen with no accessible doorway to the living room.
In the middle example, the model completely forgets to include a bathroom, an essential functional space in residential design.
In the right example, the balcony is placed in the top-right corner and adjacent to the kitchen, yielding an undersized and implausible balcony.
A more reasonable adjacency would position it in the bottom-right area next to the bedroom.
In contrast, the full model CE2EPlan (w/ Topo) equipped with the GATs consistently learns correct adjacency patterns and produces more structurally coherent and functionally plausible layouts.

\subsection{Discussion} \label{subsec:ex_discussion}

\subsubsection{Enhance output diversity}

To enhance the diversity of the generated layouts, we introduce a controlled stochasticity-injection mechanism during the reverse diffusion process.
At each reverse step $t$, after obtaining the predicted sample $x_{t}$, we perturb the sample with a small amount of Gaussian noise $x_{noise}\sim \mathcal{N}(0,\mathbf{I})$:
\begin{equation}
	x_{t} = x_{t} + \alpha(t/T)x_{noise}
	\label{eq:noise_injection}
\end{equation}
where $T$ is the total number of diffusion steps and $\alpha$ is set to $0.1$.
This timestep-dependent schedule injects relatively stronger noise at earlier time steps to encourage exploration, while gradually reducing the magnitude as $t\rightarrow0$ to maintain the geometric plausibility of the final layouts.
Subsequently, we still apply the multi-condition masking mechanism after each noisy update, ensuring that the injected stochasticity only affects the free (non-masked) parameters, while all user-specified conditions remain strictly enforced.
	
We quantitatively evaluate the output diversity of our CE2EPlan (w/o Noise) and the enhanced variant with noise injection during sampling (w/ Noise), as shown in Table~\ref{tab:quantitative_diversity}.
We adopt the metric $\text{Diversity}_\text{avg}$ introduced by~\cite{Wang-25} to measure the variability across multiple generated layouts:
\begin{equation}
	\text{Diversity}_\text{avg} = \left[ \mathbb{E}_{s \in S} \mathbb{E}_{1 \leq i < j \leq K} \text{IoU}_{s,r}(V_i,V_j) \right]_{r \in R}
	\label{eq:diversity_avg}
\end{equation}
	
For each test sample ($s$) from the test dataset containing $12K$ examples ($S$), we let the model generate five floor plan variants ($V_{i=1, \dots, K, K = 5}$) for each interaction mode.
We measure the similarity between any pair of variants ($V_i$ \& $V_j$) by computing the Intersection over Union (IoU) for the total area of each ($r$) of the six room categories ($R$), including the living room ($R_{liv}$), bedroom ($R_{bed}$), kitchen ($R_{kit}$), bathroom ($R_{bat}$), balcony ($R_{bal}$), and storage ($R_{sto}$).
For each sample ($s$), we compute the mean IoU across all variant pairs, and then average the results over the entire dataset.
The $\text{Diversity}_\text{avg}$ ranges from $0$ to $1$.
A lower value indicates a smaller IoU between the generated results, implying greater output diversity of the model.

Across $\text{Mode}_{\text{auto}}$ and $\text{Mode}_{\text{t}}$, CE2EPlan exhibits meaningful output diversity.
Although $\text{Mode}_{\text{t}}$ introduces room types as input, the design space remains sufficiently large, and the proposed noise injection mechanism effectively enhances the variability of the generated layouts.
In $\text{Mode}_{\text{t}\&\text{l}}$, however, both room types and room locations are fixed.
Under such strong constraints, the feasible design space becomes extremely limited, even for human designers.
As a result, CE2EPlan naturally produces only minor variations, and the noise injection mechanism offers little additional diversity.
We also observe an increase in FID when noise is injected, indicating a slight deviation from the data distribution of ground truths (GT).
This reflects the inherent trade-off between sample diversity and fidelity.
In practice, users may adjust the noise scale $\alpha$ in Equation~(\ref{eq:noise_injection}) to balance generation quality and diversity according to their design requirements.

\begin{table}[t]
	\centering
	\caption{Quantitative evaluation of the output diversity of our CE2EPlan (w/o Noise) and the enhanced variant with noise injection (w/ Noise) on the areas of six room categories, including the living room ($\text{R}_\text{liv}$), bedroom ($\text{R}_\text{bed}$), kitchen ($\text{R}_\text{kit}$), bathroom ($\text{R}_\text{bat}$), balcony ($\text{R}_\text{bal}$), and storage ($\text{R}_\text{sto}$). The $\text{Diversity}_\text{avg}$ ranges from $0$ to $1$, where lower scores indicate higher output diversity.}
	\label{tab:quantitative_diversity}
	\resizebox{\linewidth}{!}{\begin{tabular}{|l|c|c|cccccc|}
			\hline
			& Interaction &  & \multicolumn{6}{c|}{$\text{Diversity}_\text{avg}(\downarrow)$}\\
			CE2EPlan & Mode & FID $(\downarrow)$ & $\text{R}_\text{liv}$  & $\text{R}_\text{bed}$  & $\text{R}_\text{kit}$ & $\text{R}_\text{bat}$ & $\text{R}_\text{bal}$ & $\text{R}_\text{sto}$  \\
			\hline
			w/o Noise & $\text{Mode}_{\text{auto}}$ & \textbf{1.08} & 0.55 & 0.49 & 0.21  & 0.16  & 0.32 & \textbf{0.00} \\
			w/ Noise & $\text{Mode}_{\text{auto}}$ & 1.89 & \textbf{0.51} & \textbf{0.45} & \textbf{0.17}  & \textbf{0.12}  & \textbf{0.24} & \textbf{0.00} \\
			\hline
			w/o Noise & $\text{Mode}_{\text{t}}$ 	& \textbf{0.99}    & 0.57    & 0.53    & 0.27    & 0.18    & 0.39    & \textbf{0.00} \\
			w/ Noise & $\text{Mode}_{\text{t}}$ & 1.12    & \textbf{0.53}    & \textbf{0.48}    & \textbf{0.21}    & \textbf{0.14}    & \textbf{0.33}    & \textbf{0.00} \\
			\hline
			w/o Noise & $\text{Mode}_{\text{t}\&\text{l}}$ 	& \textbf{0.24}    & \textbf{0.93}    & \textbf{0.92}    & 0.83    & 0.79    & \textbf{0.69}    &\textbf{0.02} \\
			w/ Noise & $\text{Mode}_{\text{t}\&\text{l}}$ & 0.25    & \textbf{0.93}    & \textbf{0.92}    & \textbf{0.82}    & \textbf{0.78}    & \textbf{0.69}    &\textbf{0.02} \\
			\hline
	\end{tabular}}
\end{table}

\begin{figure}[t]
	\centering
	\includegraphics[width=\linewidth]{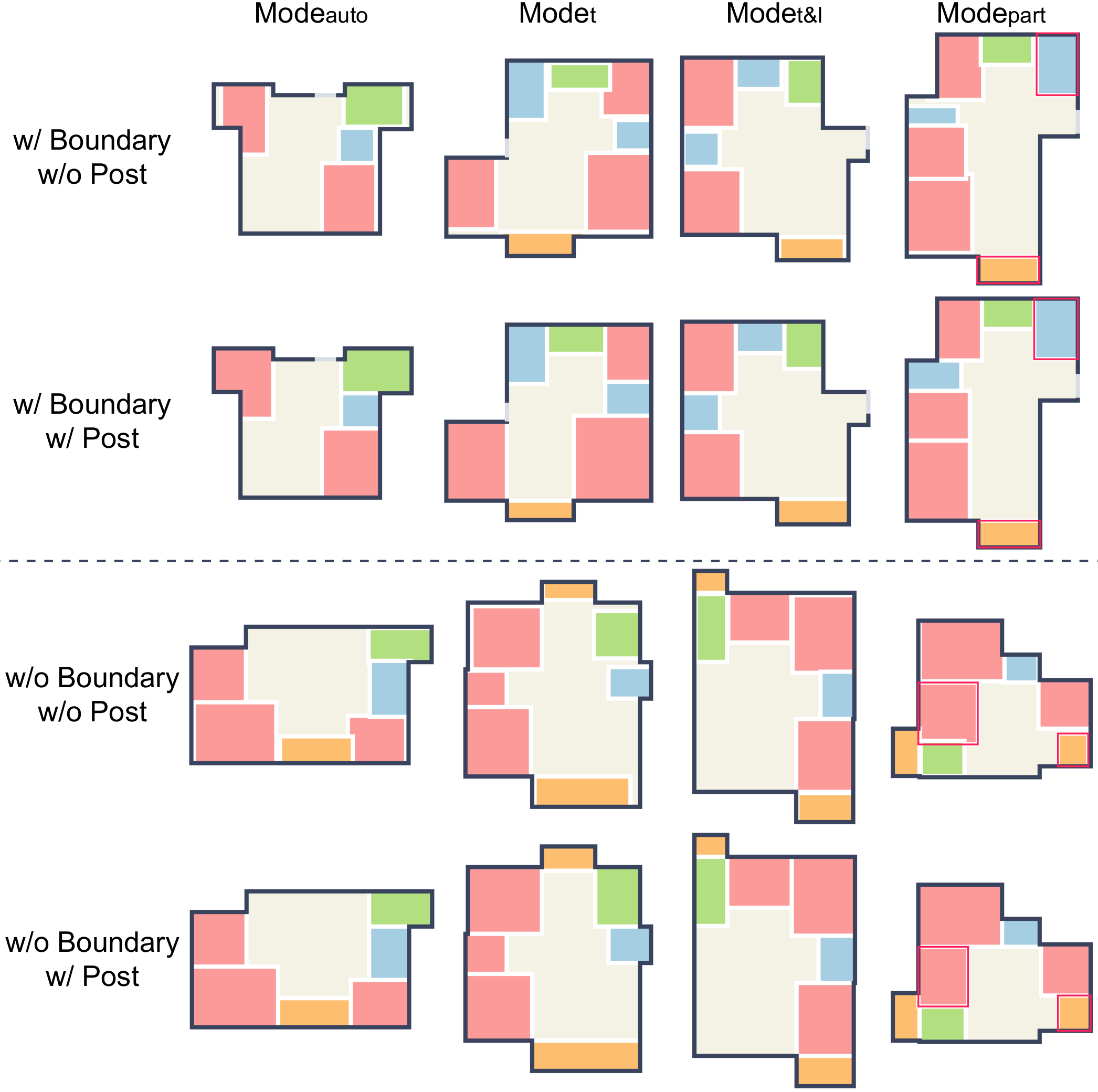}
	\caption{Some examples of the post-processing corrections (w/ Post \& w/o Post) for the model-generated results in different application scenarios (w/ Boundary \& w/o Boundary) and user interaction modes, i.e., without user input ($\text{Mode}_{\text{auto}}$), using room type input ($\text{Mode}_{\text{t}}$), using both room type \& location input ($\text{Mode}_{\text{t}\&\text{l}}$), and partial input (highlighted with red boxes, $\text{Mode}_{\text{part}}$).}
	\label{fig:ex_quali_proc}
\end{figure}

\subsubsection{Data representation and post-processing}

In this work, we use the bounding boxes to represent rooms and map them to a continuous space.
Although we design a geometry-enhanced module to improve the geometric consistency of the results generated by the model, gaps may still exist between the generated room boxes.
To address this, a simple post-processing step, derived from~\cite{Hu-20}, is required to fully eliminate these gaps.
Some examples of the post-processing corrections for the model-generated results in different application scenarios and user interaction modes are shown in Figure~\ref{fig:ex_quali_proc}.

Post-processing is a standard component in most recent automatic floor plan generation pipelines.
Different methods use different strategies depending on their data representation and pipeline: RPLAN~\cite{Wu-19} and ActFloor-GAN~\cite{Wang-21} focus on interior wall alignment, WallPlan~\cite{Sun-22} integrates wall refinement into each iteration step, iPLAN~\cite{He-22} employs a learning-based refinement module, Graph2Plan~\cite{Hu-20} and DiffPlanner~\cite{Wang-25} apply room-box alignment.
Although the post-processing step is simple and effective, in future work, we still aim to develop strategies to further reduce the need for post-processing, allowing the model to directly generate perfectly aligned floor plans.

The representation of axis-aligned bounding boxes inherently restricts the output of model to rectangular, Manhattan World layouts.
This inductive bias simplifies the learning problem but limits the geometric expressiveness of the generated designs.
Specifically, it prevents the model from producing L-shaped rooms, slanted walls, or other non-rectangular geometries commonly found in real-world architecture.
To partially mitigate this limitation and offer users more flexible room shapes, we introduce a simple post-processing merge algorithm that allows the construction of non-rectangular room geometries.
The algorithm first identifies adjacent rooms of the same category using the computed room adjacency graph.
Pairs of rooms whose shared-wall length exceeds a predefined threshold are merged into a single polygonal room.
To avoid excessively large or unrealistic merged regions, each room is allowed to participate in at most one merge operation.

\begin{figure}[t]
	\centering
	\includegraphics[width=\linewidth]{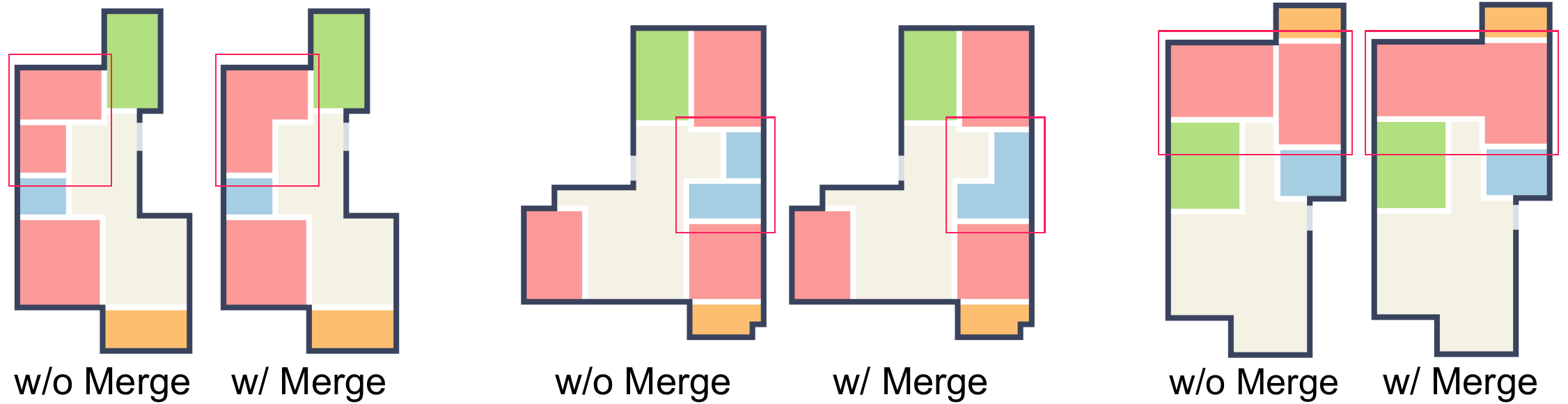}
	\caption{Some examples of L-shaped rooms (highlighted with red boxes) produced by the merge algorithm (w/ Merge) under the original room bounding boxes (w/o Merge) generated by our CE2EPlan.}
	\label{fig:ex_quali_merge}
\end{figure}

As shown in Figure~\ref{fig:ex_quali_merge}, this merge procedure enables CE2EPlan to generate L-shaped rooms (w/ Merge) that are not achievable under the original bounding-box representation (w/o Merge).
However, applying this merge step increases the FID score (from $1.09$ (w/o Merge) to $2.41$ (w/ Merge) on $\text{Mode}_{\text{auto}}$), which is likely due to the strong bias of axis-aligned rooms present in the dataset and the tendency of merged rooms to become larger than typical rooms in the data distribution of GT.
Nevertheless, the merge module provides users with an optional mechanism to obtain non-rectangular layouts when desired.
In future work, we aim to develop more principled merge strategies and alternative shape representations beyond bounding boxes, which can naturally support non-Manhattan geometries, e.g., slanted walls and curved boundaries.
Such improvements would enable significantly richer and more realistic floor plan generation capabilities.

\subsubsection{Usability of the input modalities}
While our multi-condition masking mechanism effectively consolidates various control points from traditional pipelines, the input modalities may not fully align with the workflows of professional designers.
In future work, we aim to extend CE2EPlan toward more user-friendly and semantically meaningful forms of interaction.
Potential directions include enabling users to specify high-level design intents, e.g., ``a family home for four'' or ``a compact studio apartment'', incorporating absolute dimensional or site constraints to accommodate scenarios ranging from small apartments to large villas as well as larger-scale layout tasks such as residential community planning~\cite{wang2025clodreco}, and allowing reference images or sketches as input conditions.
Such high-level inputs could be internally translated by the model into low-level constraints, including room types, counts, and approximate spatial zones, ultimately making the system more accessible and powerful for non-expert users and initial design exploration.



\section{Conclusion}

In this work, we proposed CE2EPlan, a novel controllable topology- and geometry-enhanced diffusion model for end-to-end floor plan generation.
Unlike previous approaches based on the multi-step pipeline, CE2EPlan removes the dependency on predefined solution paths and intermediate representations, instead learning comprehensive design principles from the existing data and generating high-quality floor plans directly from input in an end-to-end manner.

Extensive experiments validate that, through its novel integration of a diffusion model as the backbone, a multi-condition masking mechanism, a topology-enhanced module using GATransformer as the noise predictor, and a geometry-enhanced module, CE2EPlan significantly improves the topological coherence, geometric accuracy, and overall aesthetic quality of the generated floor plans.
Importantly, requiring only one training process, our CE2EPlan provides exceptional user control with multiple interactive modes, and inherently supports genuine output diversity, delivering multiple distinct floor plans from identical inputs without external adjustments.

In general, our approach represents a significant shift in automated floor plan generation, aligning more closely with the diverse, adaptive workflows of human designers.
By eliminating strict intermediate steps, we allow AI design tools to explore a broader solution space, ultimately enhancing both practicality and creativity in architectural layout planning.
We hope that CE2EPlan paves the way for future research in end-to-end AI-driven architectural design, fostering more intuitive, adaptable, and efficient generative models in this field.

\bibliographystyle{abbrv}
\bibliography{refs}

\begin{thebibliography}{10}

\bibitem{Arvin-02}
S.~A. Arvin and D.~H. House.
\newblock Modeling architectural design objectives in physically based space
  planning.
\newblock {\em Automation in Construction}, 11(2):213--225, 2002.

\bibitem{Bao-13}
F.~Bao, D.-M. Yan, N.~J. Mitra, and P.~Wonka.
\newblock Generating and exploring good building layouts.
\newblock {\em ACM Transactions on Graphics}, 32(4):122:1--122:10, 2013.

\bibitem{Chaillou-20}
S.~Chaillou.
\newblock {ArchiGAN}: Artificial intelligence x architecture.
\newblock {\em Architectural Intelligence}, pages 117--127, 2020.

\bibitem{chen2024lace}
J.~Chen, R.~Zhang, Y.~Zhou, R.~Jain, Z.~Xu, R.~Rossi, and C.~Chen.
\newblock Towards aligned layout generation via diffusion model with aesthetic
  constraints.
\newblock In {\em International Conference on Learning Representations}, pages
  1--12, 2024.

\bibitem{fu2024plannet}
Q.~Fu, S.~He, X.~Li, and H.~Fu.
\newblock {PlanNet}: A generative model for component-based plan synthesis.
\newblock {\em IEEE Transactions on Visualization and Computer Graphics},
  30(8):4739--4751, 2024.

\bibitem{He-22}
F.~He, Y.~Huang, and H.~Wang.
\newblock {iPLAN}: Interactive and procedural layout planning.
\newblock In {\em Proceedings of IEEE Conference on Computer Vision and Pattern
  Recognition (CVPR)}, pages 7793--7802, 2022.

\bibitem{Heusel-17}
M.~Heusel, H.~Ramsauer, T.~Unterthiner, B.~Nessler, and S.~Hochreiter.
\newblock {GANs} trained by a two time-scale update rule converge to a local
  nash equilibrium.
\newblock In {\em Advances in Neural Information Processing Systems}, pages
  6629--6640, 2017.

\bibitem{Ho-20}
J.~Ho, A.~Jain, and P.~Abbeel.
\newblock Denoising diffusion probabilistic models.
\newblock In {\em Advances in Neural Information Processing Systems}, pages
  6840--6851, 2020.

\bibitem{Hu-20}
R.~Hu, Z.~Huang, Y.~Tang, O.~van Kaick, H.~Zhang, and H.~Huang.
\newblock {Graph2Plan}: Learning floorplan generation from layout graphs.
\newblock {\em ACM Transactions on Graphics}, 39(4):118:1--118:14, 2020.

\bibitem{inoue2023layoutdm}
N.~Inoue, K.~Kikuchi, E.~Simo-Serra, M.~Otani, and K.~Yamaguchi.
\newblock {LayoutDM}: Discrete diffusion model for controllable layout
  generation.
\newblock In {\em Proceedings of the IEEE Conference on Computer Vision and
  Pattern Recognition (CVPR)}, pages 10167--10176, 2023.

\bibitem{li2022diffusion}
X.~L. Li, J.~Thickstun, I.~Gulrajani, P.~Liang, and T.~B. Hashimoto.
\newblock {Diffusion-LM} improves controllable text generation.
\newblock In {\em Advances in neural information processing systems}, pages
  4328--4343, 2022.

\bibitem{luo2021diffusion}
S.~Luo and W.~Hu.
\newblock Diffusion probabilistic models for {3D} point cloud generation.
\newblock In {\em Proceedings of the IEEE Conference on Computer Vision and
  Pattern Recognition (CVPR)}, pages 2837--2845, 2021.

\bibitem{lyu2021conditional}
Z.~Lyu, Z.~Kong, X.~Xu, L.~Pan, and D.~Lin.
\newblock A conditional point diffusion-refinement paradigm for {3D} point
  cloud completion.
\newblock In {\em International Conference on Learning Representations}, pages
  1--11, 2022.

\bibitem{Merrell-10}
P.~Merrell, E.~Schkufza, and V.~Koltun.
\newblock Computer-generated residential building layouts.
\newblock {\em ACM Transactions on Graphics}, 29(6):181:1--181:12, 2010.

\bibitem{Michalek-02}
J.~Michalek, R.~Choudhary, and P.~Papalambros.
\newblock Architectural layout design optimization.
\newblock {\em Engineering Optimization}, 34(5):461--484, 2002.

\bibitem{Nauata-20}
N.~Nauata, K.-H. Chang, C.-Y. Cheng, G.~Mori, and Y.~Furukawa.
\newblock {House-GAN}: Relational generative adversarial networks for
  graph-constrained house layout generation.
\newblock In {\em Proceedings of European Conference on Computer Vision
  (ECCV)}, pages 162--177, 2020.

\bibitem{Nauata-21}
N.~Nauata, S.~Hosseini, K.-H. Chang, H.~Chu, C.-Y. Cheng, and Y.~Furukawa.
\newblock {House-GAN++}: Generative adversarial layout refinement network
  towards intelligent computational agent for professional architects.
\newblock In {\em Proceedings of IEEE Conference on Computer Vision and Pattern
  Recognition (CVPR)}, pages 13632--13641, 2021.

\bibitem{qiu2025llmbased}
Z.~Qiu, J.~Liu, Y.~Wu, P.~Liu, H.~Qi, H.~Liang, and Y.~Xia.
\newblock {LLM-based} framework for automated and customized floor plan design.
\newblock {\em Automation in Construction}, 180(106512):1--25, 2025.

\bibitem{ramesh2022hierarchical}
A.~Ramesh, P.~Dhariwal, A.~Nichol, C.~Chu, and M.~Chen.
\newblock Hierarchical text-conditional image generation with {CLIP} latents.
\newblock {\em arXiv preprint arXiv:2204.06125}, 2022.

\bibitem{Rodrigues-13}
E.~Rodrigues, A.~R. Gaspar, and A.~Gomes.
\newblock An approach to the multi-level space allocation problem in
  architecture using a hybrid evolutionary technique.
\newblock {\em Automation in Construction}, 35:482--498, 2013.

\bibitem{Ronneberger-15}
O.~Ronneberger, P.~Fischer, and T.~Brox.
\newblock {U-Net}: Convolutional networks for biomedical image segmentation.
\newblock In {\em International Conference on Medical Image Computing and
  Computer-Assisted Intervention (MICCAI)}, pages 234--241, 2015.

\bibitem{sanfeliu1983GED}
A.~Sanfeliu and K.-S. Fu.
\newblock A distance measure between attributed relational graphs for pattern
  recognition.
\newblock {\em IEEE Transactions on Systems, Man, and Cybernetics},
  SMC-13(3):353--362, 1983.

\bibitem{Shabani-23}
M.~A. Shabani, S.~Hosseini, and Y.~Furukawa.
\newblock {HouseDiffusion}: Vector floorplan generation via a diffusion model
  with discrete and continuous denoising.
\newblock In {\em Proceedings of IEEE Conference on Computer Vision and Pattern
  Recognition (CVPR)}, pages 5466--5475, 2023.

\bibitem{Shneiderman-20}
B.~Shneiderman.
\newblock Human-centered artificial intelligence: Reliable, safe \&
  trustworthy.
\newblock {\em International Journal of Human--Computer Interaction},
  36(6):495--504, 2020.

\bibitem{song2020ddim}
J.~Song, C.~Meng, and S.~Ermon.
\newblock Denoising diffusion implicit models.
\newblock In {\em International Conference on Learning Representations}, pages
  1--12, 2021.

\bibitem{strudel2022self}
R.~Strudel, C.~Tallec, F.~Altch{\'e}, Y.~Du, Y.~Ganin, A.~Mensch, W.~Grathwohl,
  N.~Savinov, S.~Dieleman, L.~Sifre, and R.~Leblond.
\newblock Self-conditioned embedding diffusion for text generation.
\newblock {\em arXiv preprint arXiv:2211.04236}, 2022.

\bibitem{Sun-22}
J.~Sun, W.~Wu, L.~Liu, W.~Min, G.~Zhang, and L.~Zheng.
\newblock {WallPlan}: Synthesizing floorplans by learning to generate wall
  graphs.
\newblock {\em ACM Transactions on Graphics}, 41(4):92:1--92:14, 2022.

\bibitem{Vaswani-17}
A.~Vaswani, N.~Shazeer, N.~Parmar, J.~Uszkoreit, L.~Jones, A.~N. Gomez,
  {\L}.~Kaiser, and I.~Polosukhin.
\newblock Attention is all you need.
\newblock In {\em Advances in Neural Information Processing Systems}, pages
  6000--6010, 2017.

\bibitem{Velickovic-18-GAT}
P.~Veli{\v{c}}kovi{\'{c}}, G.~Cucurull, A.~Casanova, A.~Romero, P.~Li{\`{o}},
  and Y.~Bengio.
\newblock Graph attention networks.
\newblock In {\em International Conference on Learning Representations}, pages
  1--12, 2018.

\bibitem{wang2025clodreco}
S.~Wang and R.~Pajarola.
\newblock A controllable generative design framework for residential
  communities with multi-scale architectural representations.
\newblock {\em Computers in Industry}, 173(104367):1--11, 2025.

\bibitem{Wang-25}
S.~Wang and R.~Pajarola.
\newblock {Eliminating rasterization}: Direct vector floor plan generation with
  {DiffPlanner}.
\newblock {\em IEEE Transactions on Visualization and Computer Graphics},
  31(10):7906--7922, 2025.

\bibitem{Wang-21}
S.~Wang, W.~Zeng, X.~Chen, Y.~Ye, Y.~Qiao, and C.-W. Fu.
\newblock {ActFloor-GAN}: Activity-guided adversarial networks for
  human-centric floorplan design.
\newblock {\em IEEE Transactions on Visualization and Computer Graphics},
  29(3):1610--1624, 2023.

\bibitem{Wu-18}
W.~Wu, L.~Fan, L.~Liu, and P.~Wonka.
\newblock {MIQP}-based layout design for building interiors.
\newblock {\em Computer Graphics Forum}, 37(2):511--521, 2018.

\bibitem{Wu-19}
W.~Wu, X.-M. Fu, R.~Tang, Y.~Wang, Y.-H. Qi, and L.~Liu.
\newblock Data-driven interior plan generation for residential buildings.
\newblock {\em ACM Transactions on Graphics}, 38(6):234:1--234:12, 2019.

\bibitem{zhang2023text}
C.~Zhang, C.~Zhang, M.~Zhang, I.~S. Kweon, and J.~Kima.
\newblock Text-to-image diffusion models in generative {AI}: A survey.
\newblock {\em arXiv preprint arXiv:2303.07909}, 2023.

\bibitem{zong2024housetune}
Z.~Zong, G.~Chen, Z.~Zhan, F.~Yu, and G.~Tan.
\newblock {HouseTune}: Two-stage floorplan generation with {LLM} assistance.
\newblock {\em arXiv preprint arXiv:2411.12279}, 2024.

\end{thebibliography}

\begin{IEEEbiography}[{\includegraphics[width=1in,height=1.25in,clip,keepaspectratio]{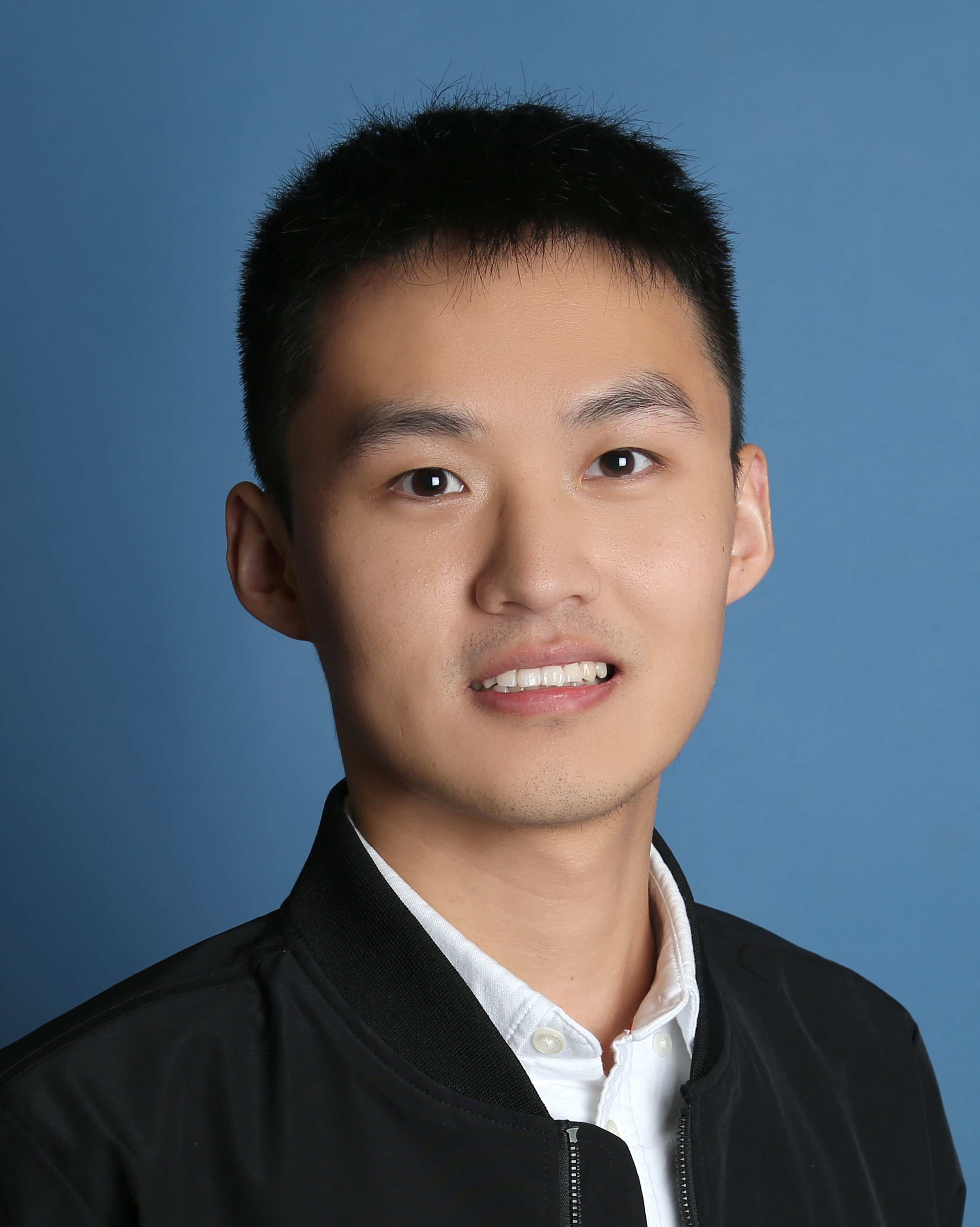}}]
{Shidong Wang} is currently working toward the PhD degree with the Visualization and MultiMedia Lab (VMML), Department of Informatics, University of Zurich.
His research interests include human-AI interaction, deep generative modeling, and computational design.
\end{IEEEbiography}

\begin{IEEEbiography}[{\includegraphics[width=1in,height=1.25in,clip,keepaspectratio]{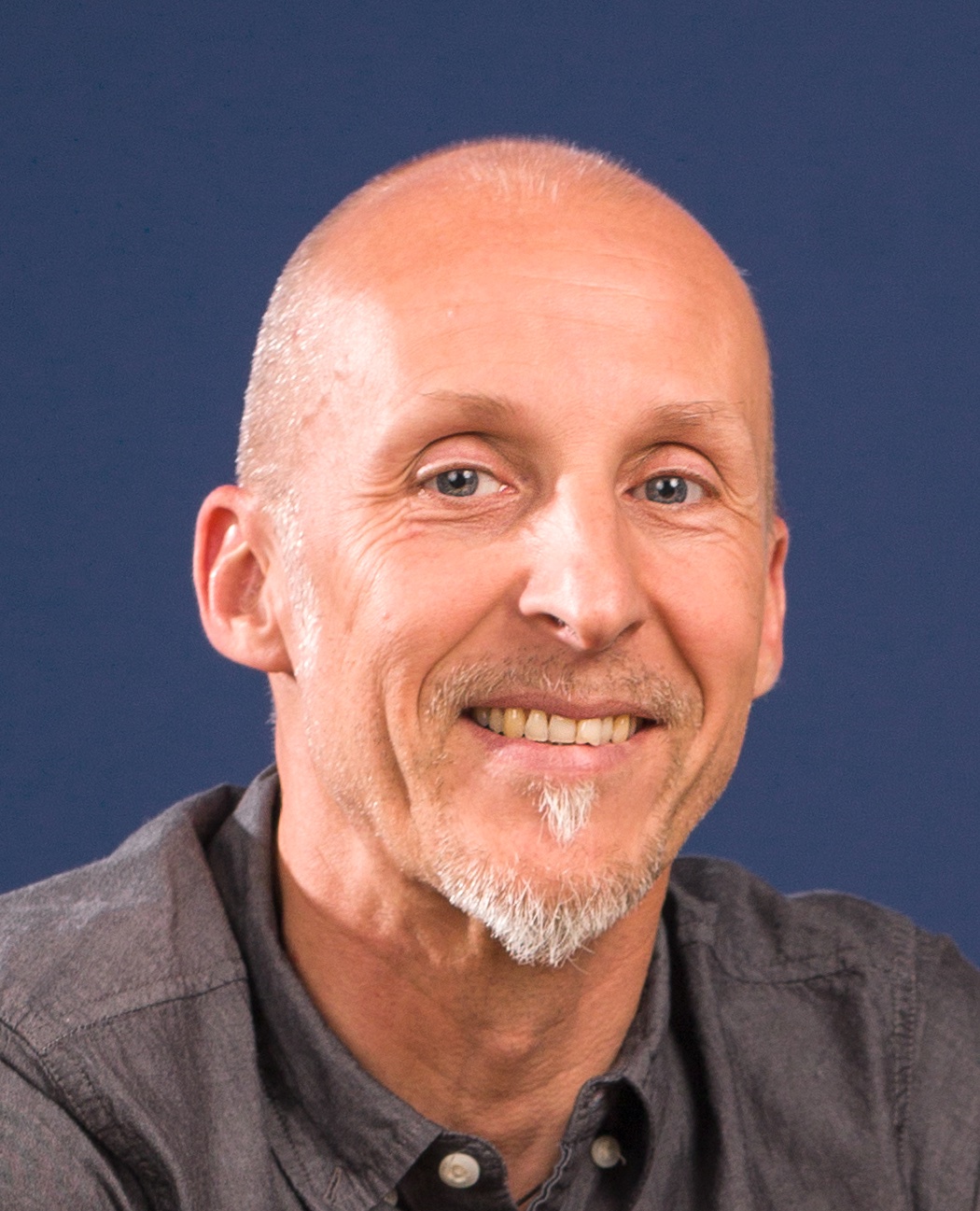}}]
{Prof. Dr. Renato Pajarola}	has been a Professor in computer science at the University of Zurich since 2005, leading the Visualization and MultiMedia Lab (VMML). He has previously been an Assistant Professor at the University of California Irvine and a Postdoc at Georgia Tech. He has received his Dipl. Inf-Ing. ETH and Dr. Sc. techn. degrees in computer science from the Swiss Federal Institute of Technology (ETH) Zurich in 1994 and 1998 respectively. He is a Fellow of the Eurographics Association and a Senior Member of both ACM and IEEE. His research interests include real-time 3D graphics, interactive data visualization, and geometry processing.
\end{IEEEbiography}

\vfill

\end{document}